\documentclass[graybox]{svmult}

\usepackage{helvet}         
\usepackage{courier}        
\usepackage{type1cm}        
\usepackage{makeidx}         
\usepackage{graphicx}        
\usepackage{multicol}        
\usepackage[bottom]{footmisc}

\makeindex             

\usepackage{pst-all}

\usepackage{latexsym}

\usepackage{amsmath}

\usepackage{amsgen}

\usepackage{amsopn}

\usepackage{amsbsy}

\usepackage{amssymb}
\usepackage{fancybox}
\usepackage{amsfonts}
\usepackage{amstext}
\usepackage{graphicx, epstopdf}

\usepackage{graphicx, epstopdf}

\newcommand\R{{\mathbb R}}

\newcommand\half{{\mbox{$\frac 12$}}}
\newcommand\eps{\varepsilon}

\newcommand{\beq}{\begin{equation}}
\newcommand{\eeq}{\end{equation}}

\newcommand{\pv}{{\bf p}}
\newcommand{\xv}{{\bf x}}

\newcommand{\x}{{\bf x}}
\newcommand{\rv}{{\mathbf r}}

\newcommand{\X}{{\bf X}}
\newtheorem{thm}{Theorem}[section]

\newtheorem{cor}{Corollary}[section]

\def\mfr#1/#2{\hbox{${\frac{#1}{#2}}$}}

\newcommand{\pot}{V_{\om,k}}

\newcommand{\rhoNm}{\rho_{N,m}}
\newcommand{\rhoNmone}{\rho_{N,m} ^{(1)}}

\newcommand{\HNm}{\mathcal{H}_{N,m}}

\newcommand{\MFmf}{\mathcal E ^{\rm MF}_{N,m}}

\newcommand{\rhoMFm}{\rho ^{\rm MF}_{N,m}}

\newcommand{\om}{\omega}

\newcommand{\intR}{\int_{\R ^2}}

\begin{document}

\title*{Topics in the Mathematical Physics\\ of Cold Bose Gases}
\author{Jakob Yngvason}
\authorrunning{Cold Bose Gases}
\institute{Jakob Yngvason \at Faculty of Physics, University of Vienna, \email{jakob.yngvason@univie.ac.at}}
%
%
\maketitle

\abstract* {In these notes of six lectures on selected topics in the theory of cold, dilute Bose gases, presented at the 5th Warsaw School of Statistical Physics in June 2013,  the following topics are discussed: 1) The concept of BEC, 2) the ground state energy of a dilute Bose gas with short range interactions, 3) Gross-Pitaevskii theory and BEC in trapped gases, 4) Bose gases in rotating traps and  quantized vortices, and 5) strongly correlated phases in the lowest Landau level generated by rapid rotation.}

\abstract {In these notes of six lectures on selected topics in the theory of cold, dilute Bose gases, presented at the 5th Warsaw School of Statistical Physics in June 2013,  the following topics are discussed: 1) The concept of BEC, 2) the ground state energy of a dilute Bose gas with short range interactions, 3) Gross-Pitaevskii theory and BEC in trapped gases, 4) Bose gases in rotating traps and  quantized vortices, and 5) strongly correlated phases in the lowest Landau level generated by rapid rotation.}

\section{Introduction}
\label{sec:1}

Ingenious experimental techniques for cooling and trapping atoms \cite{KD} have since the mid 1990's  opened a venue for studying the fascinating macroscopic quantum phenomena exhibited by of such systems \cite{PS, PiSt, Yu}.  These include Bose-Einstein condensation (BEC), superfluity, quantization of vorticity, and strong correlations produced by rapid rotations in Bose gases. Research on cold quantum gases is presently one of the most active areas of condensed matter physics. 

On the theoretical side the subject 
is, in fact, quite old, going
back to A. Einstein's paper on BEC in ideal (i.e., noninteracting) gases from 1924 \cite{E}. The theory of Bose gases with interactions began
with N.N. Bogoliubov's fundamental work of 1947 \cite{Bo}.
This was followed by a period of considerable activity in this field in
the late 1950's and early 60's. See, in particular \cite{HY,D,L2}. Due to the complexity of the quantum mechanical many-body problem, however,  mathematically rigorous results were few and are 
still hard to get. 

The challenge for mathematical physics is to 
start from a realistic many body Hamiltonian
and derive the properties of its low energy states by rigorous
mathematical analysis. Here substantial progress has been made in the past 15 years and a selection of such results is presented in these notes.

\section{The Concept of Bose-Einstein Condensation}
\label{sec:2}

The basic facts about Bose-Einstein condensation can be summarized as follows:
\begin{itemize}
\item Under normal conditions the atoms of a gas
are distributed among very many quantum states
so that every single state is only occupied by
relatively few atoms on the average.
\item In BEC a {\em single} quantum state is occupied by a
macroscopic number of atoms. This is possible if the atoms are {\em bosons}.
\item For BEC in gases that are sufficiently dilute so that interactions can be ignored in first instance,  extremely low temperatures, of the
order $10^{-8}$ K, are required. \end{itemize}
Before discussing the general concept of BEC is is appropriate to review briefly the standard textbook treatment (see, e.g., \cite{Huang}) for ideal gases.
\subsection{BEC in Ideal Bose Gases}

We consider a basis of single particle states, labelled by  
	$k=0,1,\dots$\footnote{Classically the states are points $(p,q)$ in phase space that we think of as discretized for simpler comparison with the quantum case.}  and with energies $\varepsilon_{0}\leq \varepsilon_{1}\leq 
	\varepsilon_{2}\cdots$. Classically, the $N$-particle states are  specified by $N$-tuples, 
	$(k_{1},\dots,k_{N})$, but quantum mechanically, only the  {\em occupation numbers}
	$(N_{0},N_{1},\dots)$ with $\sum_{k}N_{k}=N$ matter. For fermions, only the values $0$ or $1$ are allowed, so macroscopic occupation of one state is excluded from the outset. 
	For bosons, on the other hand,  all values $N_{k}=0,1,2,3,\dots$ are  in principle possible. 
	In the grand canonical ensemble the {\em  average occupation numbers} are given by
\beq N_{k}=\left\{ \begin{array}{lcl}
z e^{-\varepsilon_{k}/k_{\rm B}T}&{} & \text{\rm for classical Boltzmann
statistics}\\ \\ \left({z^{-1} e^{\varepsilon_{k}/k_{\rm B}T}-1}\right)^{-1}
&{}&  \text{\rm for Bose statistics} 
\end{array}
\right.\eeq
where the  fugacity $z=e^{\mu/k_{\rm B}T}\geq 0$ (with $\mu$ the chemical potential) is determined by 
 \beq \sum_{k}{} N_{k}=N.\eeq
If $\varepsilon_{0}=0$ then $z<1$ for bosons. We now consider two specific examples.
	
\subsubsection{Particles in a box}

Suppose the particles are confined in a  (large) rectangular box $\Lambda$  in $\mathbb R^d$ with  volume $|\Lambda|$.
In the {\em thermodynamic limit} where  $N,|\Lambda|\to\infty$ with  $\rho=N/|\Lambda|$ fixed, the energy values $\varepsilon_k$ scale like $|\Lambda|^{-2/d}$ and the replacement 
\beq\label{3}\sum_{k}\rightarrow \int d\varepsilon 
D(\varepsilon)\eeq 
with the {\it density of states} $D(\varepsilon)\sim |\Lambda|$ appears reasonable.
If classical statistics applies all $N_k$ are proportional to $z$, and hence
\beq \frac{N_{0}}N=\frac 1{\int_{ 0}^\infty 
e^{-\varepsilon/k_{\rm B}T}D(\varepsilon) d\varepsilon}\sim\frac 1{|\Lambda|}\to 
0\quad\text{\rm for}\quad |\Lambda|\to\infty,\eeq
i.e., there is no macroscopic occupation of the single particle ground state. 

For a Bose gas the situation is different. Here the replacement \eqref{3} can in general only be used for the excited states  and we write (again assuming $\eps_0=0$)
\begin{eqnarray}{\label{5}} N_{0}&=&\frac 1{z^{-1}-1}\\ \label{6}
N&=&{} N_{0}+\int_{\varepsilon_1}^\infty \frac{D(\varepsilon)}
{z^{-1} e^{\varepsilon/k_{\rm B}T}-1}d\varepsilon\end{eqnarray}
keeping in mind that $\eps_1\to 0$ as $|\Lambda|\to\infty$. We note also that the integral in Eq. (6) is monotonously increasing as $\eps_1\to 0$  and also if $z\to 1$ from below.
For a given $N$ and $\Lambda$, Eqs. \eqref{5} and \eqref{6} define $z=z(\rho;N)$ and hence $N_0=N_0(\rho;N)$.
There are now two cases to consider:	  
\begin{itemize}
\item[1.] BEC:
\beq\frac 1{|\Lambda|}\int_{0}^\infty \frac{D(\varepsilon)}
{e^{\varepsilon/k_{\rm B}T}-1}d\varepsilon=:\rho_{c}(T)<\infty.
\eeq
Then, if $\rho=N/|\Lambda| >\rho_{c}(T)$, $z=z(\rho;N)\to 1$ in the thermodynamic limit  and 
\beq \rho_0:=\frac{ N_{0}}{|\Lambda|}\to
\rho-\rho_{c}(T)>0
\quad\text{\rm for}\quad N\to\infty.\eeq
This is the case of BEC with a nonzero density $\rho_0$ of particles in the ground state.\smallskip
\item[2.] No BEC: \beq \label{9}\frac 1{|\Lambda|}\int_{0}^\infty \frac{D(\varepsilon)}
{e^{\varepsilon/k_{\rm B}T}-1}d\varepsilon=\infty.\eeq
Then $z(\rho;N)$ stays uniformly bounded away from 1  as $N\to\infty$, and  hence
$N_{0}/N\to 0$. Thus there is no BEC.
\end{itemize}

From these considerations it is clear that it is the behavior of $D(\varepsilon)$ for $\varepsilon\to 0$ that matters for BEC in an ideal gas.	 In $d$ space dimensions $D(\varepsilon)\sim |\Lambda| \varepsilon^{(d-2)/2}$ and
\beq\label{10} \frac 1{|\Lambda|}\int_{\varepsilon_1}^\infty \frac{D(\varepsilon)}
{z^{-1} e^{\varepsilon/k_{\rm B}T}-1}d\varepsilon\sim T^{d/2}\sum_{\ell=1}^\infty\frac{z^\ell}{\ell^{d/2}}.\eeq
Hence for  $d=3$
there is BEC  if
	\beq \rho>\rho_{c}(T)=2,612
	 \left(\frac{mk_{\rm B}T}{2\pi\hbar^2}\right)^{3/2},\eeq
	 a condition that can either be fulfilled at given $\rho$ by decreasing $T$, or at given $T$ by increasing $\rho$. The condition also be written as
	  \beq \rho^{-1/3}<\lambda_{\rm dB}=(2\pi\hbar/(mk_{\rm B}T))^{1/2}\eeq
	  where the left-hand side is the mean particle distance and $\lambda_{\rm dB}$ is the thermal de Broglie wavelength. Since the ideal gas model can only be expected to be a reasonable approximation at low densities, low temperatures are required to achieve BEC.

	  In two and one dimensions, where $D(\eps)/|\Lambda|={\rm const.}$ resp. $\sim \eps^{-1/2}$,  the integral (9) diverges at the lower boundary, or equivalently, the sum on the right-hand side of (10) diverges for $z\to 1$. Hence there is no BEC in the thermodynamic limit for $T>0$\footnote{In an ideal gas there is trivially complete BEC at $T=0$ with all particles sitting in the ground state.}. It is interesting to note, however, that the divergence of the integral (9) is only logarithmic in $|\Lambda|$ since the lower integral boundary is $\eps_1\sim |\Lambda|^{-2/d}$. Hence, also for $d=2$ and $d=1$, there is BEC in the generalized sense that $N_0/N$ stays bounded away from zero in a {\em modified thermodynamic limit}  where $|\Lambda|\sim N/\log N$ and hence $\rho\sim \log N$ for $N\to \infty$. This shows in particular that for a mathematically unambiguous definition of BEC it is important to specify how the parameters of the problem depend on $N$ as $N\to\infty$.
	  
	\subsubsection{Inhomogeneous gas in a trap}
	  
	 Consider next a trap with a quadratic external potential  in $\mathbb R^d$
	  \beq V(\xv)\sim \omega^2|\xv|^2.\eeq
	  Here the density of states is
\beq D(\varepsilon)\sim\omega^{-d}\varepsilon^{(d-1)}.\eeq
A natural `thermodynamic limit' is defined by keeping  $N \omega^{d}$, and hence $D(\eps)/N$, fixed as $N\to\infty$.

Then 
\beq \frac 1N\int {D}(\varepsilon)\frac1{z^{-1} 
e^{\beta\varepsilon}-1} d\varepsilon\sim 
T^{d}\sum_{\ell=1}^\infty\frac{z^\ell}{\ell^{d}}.\eeq
Hence there is BEC for $d$=2 and $d$=3 in this limit but not for for $d=1$. 
In three dimension the precise condition for BEC is
\beq N\omega^3> 1,21 (k_{\rm B}T/\hbar)^{3}.\eeq
Although there is no BEC in  a one dimensional trap for $N\omega$ fixed, there is generalized BEC in the sense mentioned above if $N\omega/\log N$ is kept fixed as $N\to\infty$ \cite{KD1}.

  \subsection{The Concept of BEC for an Interacting Gas}

The many-body Hamiltonian  for $N$ (spinless) Bosons with a pair interaction potential $v$ and an
external potential  $V$ has the form 
\beq\label{17}
H_N = \sum_{i=1}^{N} \left\{- \mfr{\hbar^2}/{2m}\nabla_i^2 + V(\xv_{i})\right\}+
\sum_{1 \leq i < j \leq N} v(|\xv_i - \xv_j|).
\eeq
It acts on symmetric wave functions in $L^{2}(\R^{dN})$. In the presence of interactions the energy eigenfunctions of $H_N$ are not simply symmetrized products of eigenfunctions of the one-particle operator as for a non-interacting system. Nevertheless, the concept of the average occupation, in a many-body state,  of some given single particle state makes perfect sense, and can be used for a general definition of the concept of BEC. This is most conveniently expressed through the creation and annihilation operators $\hat a^\dagger(\varphi)$ und $\hat a(\varphi)$ for a single particle space $\varphi$,  that act in the standard way on the symmetric Fock space built over the 1-particle space $L^{2}(\R^{d})$. 

If $\rho$ is any density matrix on Fock space, the average occupancy of $\varphi$ in the state corresponding to $\rho$ is
\beq N_{\varphi}={\rm trace}\left( \rho\, \hat a^{\dagger}(\varphi)\hat a(\varphi)\right)=:\langle \hat a^{\dagger}(\varphi)\hat a(\varphi)\rangle_\rho.\eeq
BEC in the many particle state $\langle\ \cdot\ \rangle_\rho$ 
means that the average occupancy is $O(N)$ for some 1-particle state $\varphi$, 
 more precisely, for some $c>0$
\beq N_{\varphi}/N\geq c>0\eeq
for all large $N$. Here \beq  N=\sum_{i}\langle \hat a^{\dagger}(\varphi_{i})\hat a(\varphi_{i})\rangle_\rho,\eeq
with  $\{\varphi_{i}\}$ an orthonormal basis in the 1-particle space, is the average total particle number in the state $\langle\ \cdot\ \rangle_\rho$ .

It is clear that the definition is only mathematically precise if the dependence of the many-body state on $N$ is specified, and  the macroscopically occupied state $\varphi$ will also in general depend on $N$.\smallskip 

A more concrete description can be given in terms of a partial trace of the density matrix $\rho$, namely the {\em reduced
1-particle density matrix} 
\beq\rho^{(1)}(\xv,\xv')=\langle
\hat a(\xv)^{\dagger}a(\xv')\rangle_\rho.\eeq
If $\rho=|\Psi\rangle\langle \Psi|$ is a pure state given by an $N$ particle  wave function $\Psi(\x_1,\dots \x_N)$, 
then
\beq 
\rho^{(1)}(\xv,\xv')=N\int\Psi(\xv,\xv_{2},\dots\xv_{N})\bar\Psi(\xv',\xv_{2},
\dots\xv_{N})d\xv_{2}\cdots d\xv_{N}.\eeq
More generally,  $\rho^{(1)}(\xv,\xv')$ is a convex combination of such expressions. 

The   1-particle density matrix is the integral kernel of a trace class operator of trace $N$. It has a spectral decomposition
\beq 
\rho^{(1)}(\xv,\xv')=\sum_{i}N_{i}\varphi_{i}(\xv)\bar \varphi_{i}(\xv')\eeq
with eigenvalues $N_{0}\geq N_{1}\geq\dots$ and orthonormal $\varphi_{i}$.

BEC 
means that 
\beq N_{0}=O(N)\eeq
while the other $N_i$ are (in general) of lower order.\footnote{If many eigenvalues are macroscopic one speak of {\em fragmented condensation.}}
The eigenfunction $\varphi_{0}(\xv)$ of the integral kernel $\rho^{(1)}(\xv,\xv')$ to the highest eigenvalue is often referred to as
 the {\em wave function of the condensate}.
Then  $N_0|\varphi_{0}(\xv)|^2$ is the {\em spatial density} and 
$N_0|\tilde\varphi_{0}(\pv)|^2$, with $\tilde\varphi_{0}$ the Fourier transform of $\varphi_0$,  the {\em momentum density} of the condensate.

For homogeneous gases in a box $\Lambda$ the wave function of the condensate can be 
expected to be 
the constant function  $|\Lambda|^{-1/2}$.
Since 
\beq N_{0}=\int\hskip-.1cm \int\bar \varphi_{0}(\xv)\rho^{(1)}(\xv,\xv')\varphi_{0}(\xv')d\xv d\xv',\eeq 
BEC means in this case
\beq |\Lambda|^{-1}\int_\Lambda\hskip-.1cm \int_\Lambda\rho^{(1)}(\xv,\xv')d\xv d\xv'=O( 
N)\eeq 
instead of $O(1)$ if there is no BEC. 
This is called {\em Off 
Diagonal Long Range 
Order}.

In contrast to ideal gases, where there is always complete BEC in the ground state, the situation is quite different as soon as interaction is added. Here already the question of BEC in the ground state is highly nontrivial and it is only this case that will be discussed in the sequel. An essential first step in this direction is the understanding of the ground state energy.
 
\section{The Ground State Energy}
\subsection{The Scattering Length}
We consider the case $d=3$ and assume a rotationally symmetric pair interaction potential $v$ of short range between particles of mass $m$. The zero energy scattering equation for the two particle scattering in the relative coordinates is
\beq - \frac{\hbar^2}{m}\nabla^2\psi+ v\psi=0.\eeq 
Writing $\psi(\xv)=u(r)/r$ with $r=|\xv|$ this is equivalent to
\beq - \frac{\hbar^2}{m}u^{\prime\prime}(r)+v(r) u(r)=0.\eeq
For $r$ larger than the range of $v$ the solution with $u(0)=0$ has the form
\beq u(r)={\rm (const.)}(r-a)\eeq with a constant $a$ that is called the {\sl scattering length} of $v$.

Equivalently, 
\beq a=\lim_{r\to\infty}\left[r-\frac{u(r)}{u'(r)}\right]\eeq
and this is finite if $v$ decreases at least as $r^{-(3+\eps)}$ at infinity.
For $\psi(\xv)=u(r)/r$ we have outside of the range of $v$
\beq \psi(\xv)={\rm (const.)}\left(1-\frac ar\right).\eeq
If $v\geq 0$, then $0\leq a\leq$ range of $v$. 
For a hard sphere potential $a$ is equal to the radius of the sphere.

If $v$ is not positive then $a$ can be negative, and  if $- \frac{\hbar^2}{m}\nabla^2+v$ has bound states, 
$a$ can be much larger than the range of $v$.

If $v\geq 0$ the scattering length can be derived from a variational principle (see \cite{LSSY}, Appendix C):
\beq \frac{4\pi \hbar^2}m a=\inf_{\psi}\int\left\{\frac{\hbar^2}{m}|\nabla\psi|^2+|\psi|^2 v\right\}d^3\x.\eeq
where the infimum is over all differentiable $\psi$ that tend to 1 at infinity. The infimum is attained for the zero energy scattering solution. The variational principle implies in particular \cite{SR}
\beq a\leq \frac m {4\pi \hbar^2}\int v(r)\, d^3\x\eeq
and the right-hand side is the first Born approximation of $a$ for a weak potential $v$. 

For positive $v$ the scattering length determines also completely the ground state energy $E_0(2,L)$ of a pair of Bosons in a large box $\Lambda$ of side length $L\gg a$: 
\beq \label{34}E_0(2,\Lambda)\approx \frac{4\pi\hbar^2}m \frac a{L^3}.\eeq
In the Born approximation the right-hand side is just $\int v/L^3$, independent of $m$.

\subsection{The Ground State Energy of a Dilute Gas}

Consider now for {$v\geq 0$} the 
Hamiltonian of $N$ Bosons in a box $\Lambda$ of side length $L$:
\beq 
H_N^{\Lambda} = - \frac{\hbar^2}{2m} \sum_{i=1}^{N} \nabla_i ^2+ 
\sum_{1 \leq i < j \leq N} v(|\xv_i - \xv_j|)\eeq 
Its ground state energy is 
\beq E_{0}(N,L)=\inf _{\Vert\Psi\Vert=1}\langle\Psi,H_N^\Lambda\Psi\rangle \eeq
and the energy per particle in the thermodynamic limit, $N=\rho L^3\to\infty$ with $\rho$ fixed, is
\beq e_{0}(\rho):=\lim E_{0}(N,L)/N.\eeq 
This quantity is independent of the boundary conditions, but for energy bounds in a finite box one usually imposes Dirichlet or periodic conditions for the upper bound and Neumann conditions for the lower bound.

For the theory of dilute gases the  {\it low density asymptotics} of $e_{0}(\rho)$ is  of fundamental importance. Low density means here that
\beq \rho a^3\ll 1\eeq
i.e., the scattering length is much smaller than the mean particle distance $\rho^{-1/3}$. The basic formula is
\beq\label{39} e_{0}(\rho)=\hbox{$\frac{2\pi \hbar^2}{m}$}{a\rho}(1+o(1))\eeq
where the correction $o(1)$ tends to zero when $\rho a^3\to 0$.\smallskip

A {\em heuristic argument} for this formula goes as follows:

``For a dilute gas only two body scattering matters'', so we simply multiply the two-body energy \eqref{34} by the number of pairs, $N(N-1)/2$, obtaining
\beq E_{0}(N,L)\approx \frac{N(N-1)}{2}E_{0}(2,L)\approx \frac{2\pi\hbar^2}m N^2\frac a{L^3}=N\frac{2\pi\hbar^2}m a\rho.\eeq

Although this argument gives the correct leading term it is very far from a rigorous proof because the ground state can be highly correlated and  it is not legitimate to regard the pairs $N(N-1)/2$ as independent. Indeed, for $d=2$ the analogous argument gives the wrong answer: The ground state energy per particle is here
$\sim \rho|\log(\rho a^2)|^{-1}$ \cite{S, LY2}, while $E_{0}(2,L)\sim L^{-2}|\log (a^2/L^2)|^{-1}$ which multiplied by $N(N-1)/2$ would give $e_0(\rho)\sim \rho|\log(a^2/L^2)|^{-1}\to 0$ for $L\to\infty$.\smallskip

The  formula \eqref{39} for the energy of a dilute Bose gas has an interesting history and it took almost 70 years to establish the leading term rigorously. The earliest reference is a paper of W. Lenz of 1929 \cite{Len} for a gas of hard spheres, using essentially the heuristic argument above. Bogoliubov's work of 1947 \cite{Bo} was a milestone, but it was a perturbative result with $a$ replaced by its first and second Born approximations. In the 1950's and 60's several derivations were presented \cite{HY, L2}, some containing also higher order terms:
 \beq\label{41} \frac {e_{0}(\rho)}{(2\pi\hbar^2/m)a\rho}=
\left[1+{\frac{128}{15\sqrt \pi}}(\rho a^3)^{1/2}+8\left({\frac{4\pi}{ 
3}}-\sqrt 3\right)(\rho a^3)\log (\rho a^3)+O(\rho a^3)\right]\eeq
These derivations all relied on some special
assumptions about the ground state that have never been proved, or on the
selection of special terms from a perturbation series which likely diverges.  The only mathematically rigorous result in this period was due to F. Dyson \cite{D}, who in 1957 proved for a gas of hard spheres the estimates
\beq \frac{1+2 y^{1/3}}{(1-y^{1/3})^2}\geq 
\frac{e_{0}(\rho)}{(2\pi\hbar^2/m)a\rho}\geq{\frac1{10\sqrt 2}} \eeq
with $y:=4\pi a\rho^3/3$. While the upper bound gives the desired result for $\rho a^3\to 0$, the lower bound (although obtained by an ingenious reasoning) is off the mark by a factor 14.

Dyson's upper bound can be generalized to all short range interaction potentials $v\geq 0$ \cite{LSY1}. It is proved by a clever choice of a trial function, using the zero energy scattering solution for the two-body problem as an input.

An
asymptotically correct lower bound was not obtained until 1998 \cite{LY1}:
\begin{thm}[{\bf Lower bound for g.s.e.}]
\beq\label{43} \frac{e_{0}(\rho)}{(2\pi\hbar^2/m)a\rho}\geq (1-8.9\, y^{1/17}).\eeq
\end{thm}
The negative sign of the error term and the exponent 1/17 result entirely from the technique of the proof and are not believed to  reflect the true state of affairs.
In the recent paper \cite{LeeY} the error term is improved to $-O(y^{1/3}|\log y|^3)$ by a modification of the method of \cite{LY1}.

It is remarkable that the same formula, \eqref{39}, holds in two physically different regimes:

\begin{itemize}
\item[1.] `Hard potential', i.e., $v$ large within its range (in particular hard core).  The energy is here {\it mostly kinetic}, due to the bending of the wave function down to small values when two points come close together. The ground state is {\it  highly 
correlated} and simple perturbation theory is not applicable.\smallskip

\item[2.] `Soft potential', i.e., $v$ small everywhere. The energy is here {\it mostly potential}. Lowest order perturbation theory
(with the {\it  uncorrelated}, unperturbed state $\Psi_{0}=L^{-3N/2}$) 
gives 
\beq e_{0}(\rho)\approx \mfr1/2\rho\int v(\xv)d^3\xv.\eeq
This is independent of $\hbar$ and $m$ and hence cannot be the right answer, but it is in accord with the first Born approximation 
for $a$.
\end{itemize}
It is still not entirely clear if the terms in \eqref{39} beyond leading order distinguish these regimes. See \cite{ESY, YY, GS} for recent rigorous work on the second term in \eqref{41} (the ``Lee-Huang-Yang term").\smallskip

In the  analysis of the ground state energy three different length scales are involved:
\begin{itemize}
\item The scattering length $a$.
\item The mean particle distance $\rho^{-1/3}\gg a$.
\item The `healing length' $\ell_c$, defined by $1/\ell_{c}^2\sim e_{0}(\rho)$. \smallskip

For a dilute gas
\beq \ell_{c}=(a\rho)^{-1/2}=(\rho a^3)^{-1/6}\rho^{-1/3}\gg \rho^{-1/3}.\eeq 
\end{itemize}
Note that for fermions $\ell_{c}\sim \rho^{-1/3}$, i.e., much shorter than for bosons, which at low density and temperatures loose their individuality in the sense that the gas cannot be thought of as a system composed of well localized individual wave packets.

A key ingredient in the proof \cite{LY1} of the lower bound is a lemma of Dyson \cite{D}, that allows the replacement of a short range `hard' potential $v$ by a  `soft' potential $a\,U_R$ of long range $R$, at the cost of {sacrificing kinetic energy} and interactions beyond nearest neighbours. 
Dyson's Lemma, in a way,  thus transforms Regime 1 into Regime 2. Borrowing a small bit of the kinetic energy, perturbation theory applies, but one has to control the errors! For this it is necessary to divide $\Lambda$ into smaller boxed of side length $\ell$ that stays  fixed as $L\to\infty$. 
In order for this strategy to work the parameters must satisfy
\beq a\ll R\ll \rho^{-1/3}\ll \ell \ll \ell_c.\eeq 
Optimizing the choice of $R$, $\ell$ and $\varepsilon$ leads to the error factor in \eqref{43} as shown in \cite{LY1}.

\section{Gross-Pitaevskii Theory}

Consider again the $N$-body Hamiltonian \eqref{17} with an external confining potential $V$.  In this Section we choose units  so that $\hbar=1$, $m=\half$ and write
\beq 
H_N = \sum_{i=1}^{N} \left\{-  \nabla^2_i + V(\xv_{i})\right\}+
\sum_{1 \leq i < j \leq N} v(|\xv_i - \xv_j|).
\eeq
The trap potential $V$ is assumed to be nonnegative, locally bounded, 
and tend to $\infty$ for $|\xv|\to\infty$. In typical experiments with 
trapped gases $V$ can 
often be assumed to be quadratic, but this is not necessary and more general potentials will in particular become important in Sections 5 and 6. The interaction $v$ is, for the purpose of the mathematical model, supposed to be of finite range, rotationally symmetric, and nonnegative.\\

The external potential comes with a natural length scale $L_{\rm trap}=
e_V^{-1/2}$ where
$e_V$ is the spectral gap between the ground state and the first excited state of $- \nabla^2+V$. 

We are interested in the ground state properties of $H_N$, and in particular BEC,  in the {\it 
Gross-Pitaevskii (GP) limit} where $N\to\infty$ with a fixed value of the {\em GP interaction parameter}
\beq \label{48} g:= 4\pi Na/{L_{\rm trap}}=e_0(\bar \rho)/e_V\eeq
with $\bar \rho=N/{L_{\rm trap}^3}$ a measure of the mean density.
Note that since $\bar\rho a^3\sim g/N^2=O(1/N^2)$ if $g$ is fixed, the GP limit is a special case of a dilute limit.\smallskip

The GP limit can be achieved in two ways:
\begin{itemize}
\item
Keeping $a$ fixed and scaling the external potential $V$ so that $L_{\rm trap}\sim N$ (not $\sim N^{1/3}$ as in the thermodynamic limit!), i.e, writing 
\beq \label{49}V(x)=N^{-2}V_1(N^{-1}x)\eeq
 with $V_1$ fixed.\smallskip

\item Keeping $V$ fixed and taking $a\sim N^{-1}$.  The latter can be achieved by scaling the interaction potential: If $v_1$ is fixed with scattering length $a_1$, then
\beq \label{50}v(r)=N^2v_1(Nr)\eeq
has scattering length $a=a_1N^{-1}$. 
\end{itemize}\smallskip

The alternatives \eqref{49} and \eqref{50} are completely equivalent although physically it may appear more natural to regard $v$ as fixed and scale $V$. Mathematically it is usually more convenient, however, to do the opposite. The ground state energy
\beq E_0(N,a)=\inf _{\Vert\Psi\Vert=1}\langle\Psi,H_N\Psi\rangle \eeq
can then be regarded as a function of $N$ and $a=N^{-1}a_1$ for $V$ and $v_1$ fixed.

\subsection{The GP Energy Functional}

In the GP limit the essential features of the many-body ground state can be captured by 
minimizing a functional of functions on 
$\R^3$, the {\it  GP energy functional} 
\beq \mathcal E^{\rm 
GP}[\varphi]=\int_{\R^3}\left(|\nabla\varphi|^2+V|\varphi|^2+
g|\varphi|^4\right),\eeq
with the subsidiary condition $\int|\varphi|^2=1$.

The motivation for the term $g|\varphi|^4$ is the formula \eqref{39} for the energy of a dilute gas:  With $\rho(\xv)=N|\varphi(\xv)|^2$ interpreted as a local density, we have
\beq Ng\int|\varphi|^4= 4\pi a\int \rho^2,\eeq
and $4\pi a \rho(\xv)^2$ is the interaction energy per unit volume. By standard methods it can be shown that this  minimization problem has solution that is unique up to a phase factor that can be chosen so that the minimizer, denoted henceforth by $\varphi^{\rm 
GP}(\xv)$,  is positive.

The minimizer  is also the unique, nonnegative solution of the (time independent) {\it Gross-Pitaevskii equation}
\beq (- \nabla^2 + V+2g|\varphi|^2)\varphi=
\mu^{\rm GP}\,\varphi\eeq
with a Lagrange multiplier (chemical potential) $\mu^{\rm GP}$ to take care of the normalization  $\int|\varphi|^2=1$.

 The GP energy is
\beq E^{\rm GP}(g)={\mathcal E}^{\rm 
GP}[\varphi^{\rm 
GP}]=\inf\{{\mathcal E}^{\rm 
GP}[\varphi]:\, \hbox{$\int|\varphi|^2=1$}\}\eeq
and  multiplying the GP equation by $\varphi^{\rm 
GP}$ and integrating we obtain 
\beq \mu^{\rm GP} = {E}^{\rm GP}(g)+g\int|\varphi^{\rm 
GP}|^4.\eeq

The GP energy functional can be obtained {\it  formally} from the many body 
Hamiltonian by replacing $v(\xv_{i}-\xv_{j})$ with
$8\pi a\,\delta(\xv_{i}-\xv_{j})$ and making a Hartree type product ansatz for 
the many body wave function, i.e., writing
\beq \Psi(\xv_1,\dots,\xv_N)=\varphi(\xv_1)\cdots\varphi(\xv_N).\eeq
This is not a proof, however, and the true ground state is {\it not} of this form (except for $v=0$). In particular, if  $v$ has a hard core,  then $\langle\Psi, H\Psi\rangle=\infty$ for all such product wave functions. Finite energy can in this case only be obtained for functions of the form 
\beq \Psi(\xv_1,\dots,\xv_N)=\varphi(\xv_1)\cdots\varphi(\xv_N)F(\xv_1,\dots,\xv_N)\eeq 
with $F(\xv_1,\dots,\xv_N)=0$ if $|\xv_i-\xv_j|\leq a$ for a pair $i\neq j$. The upper bound on the energy is, in fact, proved by using trial functions of this form with a judiciously chosen $F$ involving the zero-energy scattering solution of the two-body problem.\smallskip

\noindent{\it Remark:}  Formally, if $v_1(r)\sim \delta(\mathbf x)$, then 
\beq v(r)=N^2v_1(Nr)=N^{-1+3\beta} v_1(N^\beta r)\eeq
independently of $\beta$. For a {\it bona fide} 3D potential $v_1$, on the other hand, the right-hand side depends on $\beta$. The GP limit corresponds to  $\beta=1$. This is the case for which the scattering length is of the same order as the range of the potential. For $0<\beta<1$ and integrable $v$ one may expect that the interaction term is $\sim (\int v)\int |\varphi(\x)|^4 \rm d\x$ , i.e., the scattering length is replaced by its first Born approximation.\footnote{See the discussion in  \cite{ESY0, Pickl} for the time dependent GP equation.}
The case $\beta=0$ is the Hartree limit, where the interaction term becomes $\int\int |\varphi(\x)|^2v(\x-\mathbf y)|\varphi(\mathbf y)|^2\mathrm d\x \mathrm d\mathbf y$. \medskip

Let now $E_0(N,a)$ denote the many body ground state energy and ${\rho^{(1)}(\xv,\xv')}$ the one-particle density matrix of the ground state. Basic results in GP theory are the following {\it  rigorous} theorems \cite{LSY1, LSe, Se5, Se6}:

\begin{thm}[{\bf Energy asymptotics}]\label{4.1}
If $N\to\infty$ with $g$ fixed (i.e., $a\sim N^{-1} L_{\rm trap}$), then
\beq\frac{E_{0}(N,a)}{NE^{\rm GP}(g)}\to1.\eeq\end{thm} 

\begin{thm}[{\bf BEC in GP limit}] \label{4.2}
If $N\to\infty$ with  $g$ 
fixed, then
\beq\label{62}\frac 1N\rho^{(1)}(\xv,\xv')\to \varphi^{\rm GP}(\xv)\varphi^{\rm 
GP}(\xv')\quad\text{in trace norm}.\eeq\end{thm} 
In other words: There is {\em complete BEC in the GP limit} and the GP minimizer is the wave function of the condensate.
\begin{cor} In the GP limit the normalized particle density in the many-body ground 
state converges to  $|\varphi^{\rm GP}(\xv)|^2$ and the normalized {\em momentum 
density} to $|\tilde\varphi^{\rm GP}(\pv)|^2$.\end{cor}

\begin{figure}[htf]
\center
\fbox{\includegraphics[width=8cm]{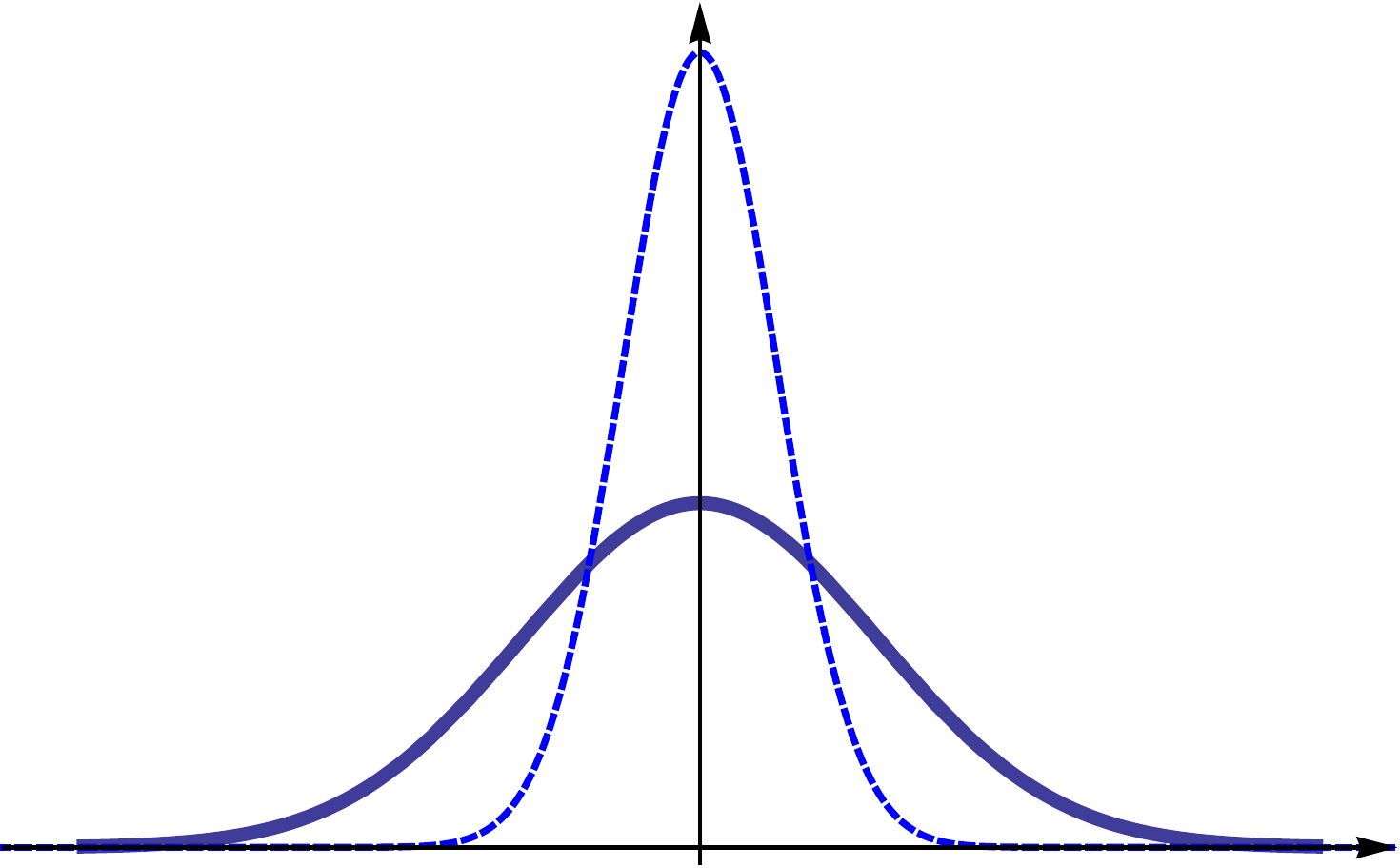}}
\caption{The GP density (fat curve) and the density without $g|\varphi|^4$ (dotted).}
\label{fig1}
\end{figure}

\subsection{The `Thomas-Fermi' approximation}

It is instructive and important to consider the properties of $\varphi^{\rm GP}$ as the interaction parameter $g$ varies, in particular the limiting case $g\gg 1$. Here it is convenient to assume  that the trap potential $V$ is a homogeneous function of some order $s>0$, i.e., $V(\lambda\xv)=\lambda^sV(\xv)$ for $\lambda>0$. 

For large $g$ salient features of the GP minimizer and energy can be estimated by the following `back of the envelope' calculation: Let {$R$} be the spatial extension of the condensate. Then $|\varphi|^2\sim R^{-3}$ and   the three terms in the GP energy functional are of the following orders of magnitude:

\beq\int|\nabla\varphi|^2\sim R^{-2},  \quad \quad
\int V|\varphi|^2\sim R^s,  \quad \quad
\quad g\int |\varphi|^4\sim gR^{-3}.
\eeq
For large $g$  the minimum of $R^{-2}+R^s+gR^{-3}$ is obtained for 
\beq R\sim g^{1/(s+3).}\eeq 
In particular for $s=2$ (quadratic trap) we obtain $R\sim g^{1/5}$.\smallskip

Note also that hat  $\int|\nabla\varphi|^2\sim R^{-2}\sim g^{-2/(s+3)}$ but the other terms of the energy are $\sim g^{s/(s+3)}$.
Hence the kinetic term becomes irrelevant for $g\gg 1$.\smallskip

To make this a little more precise we write $\xv=g^{1/(s+3)}\xv'$ and obtain
\beq
\mathcal E^{\rm GP}[\varphi]=g^{s/(s+3)}\int_{\R^3}\left(g^{-(s+2)/(s+3)}|\nabla\varphi'|^2+V|\varphi'|^2+|\varphi'|^4\right)\mathrm d^3\xv'
\eeq
with $\varphi'(\xv')=g^{3/2(s+3)}\varphi(\xv)$. Denoting $|\varphi'(\xv')|^2$ by $\rho(\xv')$  we see that in the limit $g\to\infty$ the GP functional simplifies to  
the so-called Thomas-Fermi (TF) functional\footnote{This (somewhat unfortunate)  denomination is due to a purely formal similarity with the density functional of Thomas-Fermi theory \cite{L5} for fermions.}
\beq {\mathcal E}^{\rm TF}[\rho]=\int_{\R^3}\left(V\rho+
\rho^2\right)\eeq
with normalization {$\int \rho=1$}. \smallskip

The minimizer can be displayed explicitly:
\beq\rho^{\rm TF}(\xv)=\hbox{$\frac 12$}[\mu^{\rm TF}-V(\xv)]_{+}\eeq
where $\mu^{\rm TF}$ is chosen so that the normalization condition is 
fulfilled and $[t]_+=t$ if $t\geq 0$ and zero otherwise. The corresponding {\em TF energy} is 
\beq E^{\rm TF}=\inf_\rho{\mathcal E}^{\rm TF}[\rho]={\mathcal E}^{\rm TF}[\rho^{\rm TF}]\eeq
where the infimum is over all nonnegative $\rho$ with $\int \rho=1$. Moreover,
\beq \mu^{\rm TF}=E^{\rm TF}+2\int\left(\rho^{\rm TF}\right)^2.\eeq

\begin{figure}[htf]
\center
\fbox{\includegraphics[width=8cm]{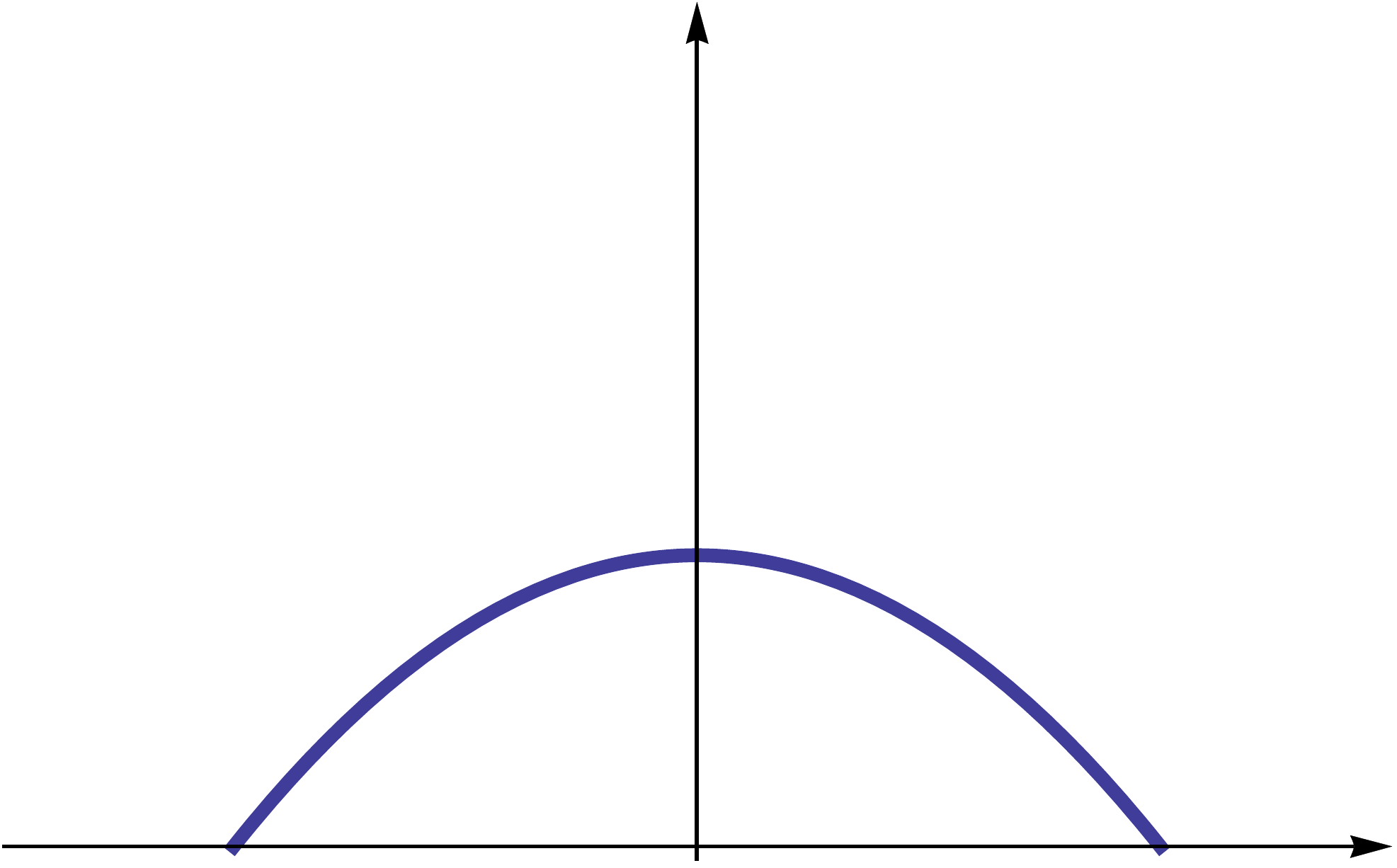}}
\caption{The TF density.}
\label{fig2}
\end{figure}

 The TF minimizer and the TF ground state energy reproduce correctly the (suitably scaled) energy and particle density of the many body ground state of the many-body problem in the $N\to\infty$, $g\to\infty$ limit under the additional hypothesis that the gas remains dilute, i.e., $a^3\bar \rho\ll 1$ where $\bar\rho$ is the average density, cf \cite{LSSY}, Sec. 6.1. But it should be noted that BEC has not yet been proved in this limit, only in the GP limit where $g$ stays fixed, or grows at most very slowly with $N$.

\subsection{On the proof of BEC in the GP Limit}

The proof of BEC  for trapped, dilute gases in \cite{LSe} has two main ingredients:

\begin{itemize}
\item
A refinement of the energy estimate \eqref{43}.  In fact,  the deviation of \eqref{43} in \cite{LY1} uses only part of the kinetic energy density, concentrated in a region in configuration space where  two particles are close together. The matching upper bound to the energy implies that the neglected part of the kinetic energy is small and this implies an $L^2$-bound on the average kinetic energy in the complementary region.

\item
An extension of 
a classical Poincar\'e inequality that estimates an $L_{p}$ 
norm of the average value of a function by an $L_{q}$ norm of its 
gradient  \cite{LSY7}.
\end{itemize}

Theorem \ref{4.2} holds for general trapping potentials $V$. The essential ideas can, however, be explained in a simplified setting, namely when the gas is confined in a box $\Lambda$ of side length $L$ with Neumann boundary conditions so the wave function of the condensate is the constant function $\varphi_0=L^{-3/2}$.
We employ the notations $\X=(\x_2,\dots,\x_N)$ and $\psi_\X(\x)=\Psi_0(\x,\X)$ with $\Psi_0$ the many-body ground state wave function.
The depletion of the condensate is
\begin{multline} \label{70}1-N_0/N=1-(NL^3)^{-1}\int_\Lambda\int_\Lambda\rho^{(1)}(\x,\x')d\x d\x'\\ =\int_{\Lambda^{N-1}} \Vert \psi_\X(\cdot)-\langle \psi_\X\rangle\Vert^2_{L^2(\Lambda)}d\X\end{multline}
where $\langle f\rangle=L^{-3}\int_\Lambda f$ denotes the average of a function $f$ over the box.\smallskip

There is a simple Poincar\'e inequality that estimates the deviation of a function from its mean value in terms of a norm of the gradient:
\beq\Vert f-\langle f\rangle\Vert^2_{L^2(\Lambda)}\leq C L^2\Vert\nabla f\Vert^2_{L^2(\Lambda)}\eeq
This inequality is a straightforward consequence of the spectral decomposition of the Neumann Laplacian in the box. A more refined inequality allows to replaces the $L_2$ norm on the right-hand side with the $L_{6/5}$ norm, and combing this with H\"older's inequality one obtains
 for arbitrary (measurable) $\Omega\subset \Lambda$ 
\beq \label{72}\Vert f-\langle f\rangle\Vert^2_{L^2(\Lambda)}\leq C_1L^2\Vert\nabla f\Vert_{L^2(\Omega)}^2+C_2|\Omega^c|^{2/3}\Vert\nabla f\Vert^2_{L^2(\Lambda)}.\eeq
where $\Omega^c$ is the complement of $\Omega$ in $\Lambda$.

  This inequality is now combined with a localization of the kinetic energy that is `hidden' in the proof \cite{LY1} of the asymptotic formula $e_0\sim \rho a (1+o(1))$ for the ground state energy:

   While the total kinetic energy per particle is of the order $\rho a$, 
 \beq \label{73}t_{\rm kin}(\Lambda)=\int_{\Lambda^{N-1}} \Vert\nabla\psi_\X\Vert^2_{L^2(\Lambda)}d\X\sim \rho a\,(1+o(1)),\eeq 
   an inspection of the proof in \cite{LY1} reveals that here is an $\Omega\subset \Lambda$ such that  $|\Omega^c|=L^3\times o(1)$ and
   \beq \label{74}t_{\rm kin}(\Omega)= \int_{\Lambda^{N-1}} \Vert\nabla\psi_\X\Vert_{L^2(\Omega)}^2d\X= \rho a \times o(1).\eeq
   Hence, if $N\to \infty$ with $g\sim Na/L$ fixed, then the combination of \eqref{70}, \eqref{72}, \eqref{73} and \eqref{74} gives
   \beq1-N_0/N\leq L^2 \rho a\times o(1)=\frac{Na}L \times o(1)\to 0.\eeq

\section{Rotating Bose Gases and Quantized Vortices}
\subsection{Quantization of Vorticity in a Superfluid}
Consider a fluid with velocity field $\mathbf v(\mathbf r)$. The circulation around a closed loop $\mathcal C$ enclosing a domain $\mathcal D$ is, by Stokes,\\
\beq \oint _{\mathcal C}\mathbf v\cdot d{\mathbf \ell}=\int_{\mathcal D}(\nabla\times\mathbf  v)\cdot\mathbf n\, {\rm d}S.\eeq
Hence nonzero circulation requires that the {\em vorticity},
\beq \nabla\times \mathbf v,\eeq
is nonzero somewhere in $\mathcal D$.
A region where $\nabla\times \mathbf v\neq 0$ is called a {\it vortex}.\smallskip

The state of a superfluid can phenomenologically be described by a complex valued function (`order parameter') $\psi(\x)=e^{\mathrm i \varphi(\x)}|\psi(\x)|$ satisfying a nonlinear Schr\"odinger Equation (the time dependent Gross-Pitaevskii equation)
\beq{ \mathrm i}\hbar \partial \psi/\partial t=-\frac{\hbar ^2}{2m}\nabla^2\psi+F(\psi)\psi\eeq
with $F(\psi)$ real valued.\footnote{In a rotating frame $F$ is not real valued, cf. Eq. \eqref{89}, but this leads only to the addition of the constant term $-2\mathbf \Omega_{\rm rot}$ to the vorticity.} The modulus squared, $\rho(\x)=|\psi(\x)|^2$, corresponds to the density of the superfluid  while the phase determines the velocity of the flow:
\beq \label{79} \mathbf v=\frac{\hbar}m\nabla \varphi.\eeq
To see this we note that just like for the linear Schr\"odinger equation, the nonlinear Schr\"odinger equation implies the continuity equation
\beq \partial\rho/\partial t+\nabla \cdot\mathbf j=0\eeq
with the current density
\beq{\mathbf j}=\frac \hbar{2m\rm i}(\bar\psi\nabla \psi-\psi\nabla\bar\psi).\eeq
Writing $\mathbf j(\x)=\rho(\x)\mathbf v(\x)$
one obtains \eqref{79}.\smallskip

Since $\psi$ is single valued we have $\oint_{\mathcal C}\nabla \varphi\cdot d\mathbf\ell=n\,2\pi$ with $n\in\mathbb Z$,
so
\beq \oint_{\mathcal C}\mathbf v\cdot d\mathbf\ell=n\frac hm\,.\eeq
This means that vorticity in a superfluid is quantized in units of $h/m$, as noted by L. Onsager in 1949 \cite{O}.

On the other hand, where the phase is nonsingular, i.e., where $|\psi(\rv)|\neq 0$, we have
\beq\nabla\times \mathbf v=0.\eeq
Generically, the complex function $\psi$ vanishes at most on one dimensional curves in $\mathbb R^3$ (points in $\mathbb R^2$) and by \eqref{79} the flow is irrotational outside these vortex lines (resp. points). In contrast, rigid rotation with $\mathbf v(\x)={\mathbf \omega}\times \x$ has $\nabla\times \mathbf v=2{\mathbf \omega}$ everywhere.

\subsection{The Many-body Hamiltonian in a Rotating Frame}

We now consider $N$ spinless
bosons in trap potential, $V$, and with a pair interaction potential, $v$, like in Section 4, but  in addition we impose a uniform rotation with angular velocity $\mathbf \Omega_{\rm rot}$ on the system, including the trap potential.  The Hamiltonian {\em in the rotating frame} is
\begin{equation}\label{84}
H_N = 
\sum_{j=1}^{N} \left(- \half \nabla^2_j +V(\x_{j})-{\mathbf L}_j\cdot {{\mathbf \Omega}_{\rm rot}}\right)+
\sum_{1 \leq i < j \leq N} v(|\x_i - \x_j|),
\end{equation}
Here  ${\mathbf L}_j=-\mathrm i \mathbf x_j\times\nabla_j$ is the angular momentum of the $j$th particle.
The Hamiltonian can alternatively  be written in the \lq magnetic' form\footnote{Here and in the sequel we choose units so that the mass $m$ is 1 rather than $\half$ as in Section 4.}
\begin{equation}\label{85}
H_N= 
\sum_{j=1}^{N} \left\{\half (\mathrm i\nabla_j+{\mathbf A}({\x}_j))^2+V(\x_{j})-\hbox{$\frac 12$}\Omega_{\rm rot}^2r_j^2\right\}+
\sum_{1 \leq i < j \leq N} v(|\x_i - \x_j|)
\end{equation}
with the vector potential 
\beq\label{vecpot} {\mathbf A}(\x)={{\mathbf \Omega}_{\rm rot}}\times \x=\Omega_{\rm rot} r\, \mathbf e_\theta\eeq
where $r=(x_1^2+x_2^2)^{1/2}$ denotes the distance from the rotation axis and $\mathbf e_\theta$ the unit vector in the angular direction. This way of writing the Hamiltonian corresponds to the splitting of the rotational effects into Coriolis and centrifugal forces. In the magnetic analogy the vector potential corresponds to a magnetic field $\mathbf B=\nabla \times \mathbf A
=2\mathbf \Omega_{\rm rot}$. 

A notable feature of the hamiltonian \eqref{84} is that, in contrast to the non-rotating case,  the bosonic ground state is in general not the same as the absolute ground state (i.e., the ground state without symmetry requirement), and it need not be unique \cite{Se4}.


\subsection{Harmonic vs.\ Anharmonic Traps}

If $V$ is a {harmonic} oscillator potential in the direction $\perp$ to ${\mathbf \Omega}_{\rm rot}$, i.e., 
\beq V(\x)=\half\Omega_{\rm trap}r^2+V^{\parallel}(x_3)\eeq
with $x_3$ the coordinate in the direction of the  axis of rotation, then stability requires
$\Omega_{\rm rot}< \Omega_{\rm trap}$. {\it Rapid rotation} means here that
\beq \Omega_{\rm rot}\to\Omega_{\rm trap}\eeq
from below. On the other hand, if  $V$ is anharmonic and increases faster than quadratically in the directions $\perp$ to $\mathbf\Omega_{\rm rot}$, e.g. $V(\x)\sim \left(r^s+V^{\parallel}(x_3)\right)$ with $s>2$, then $\Omega_{\rm rot}$ can in principle be as large as one pleases and `rapid rotation' means simply that $\Omega_{\rm rot}\to\infty$. \smallskip

These two varieties of rapid rotation turn out to differ both physically and mathematically. The former, that we shall discuss in more detail in Section 6, leads to an effective many-body Hamiltonian in the {\it lowest Landau level} of the magnetic kinetic energy term in \eqref{85} and bosonic analogues of the {\it Fractional Quantum Hall Effect} (see \cite{Fe1, Vie,Co, LeS}). On the other hand, in the case of rapid rotation in an anharmonic trap it is usually sufficient to employ Gross-Pitaevskii (GP) theory for an effective description. We remark, however, that a small anharmonic term, appropriately tuned, also leads to interesting modifications of the Quantum Hall states of harmonic traps  \cite{RSY, RSY2}. This will be discussed further in Sections 6.3 and 6.4.


\subsection{The Gross-Pitaevskii Limit Theorem with Rotation}

The following extensions of the  GP limit theorems \ref{4.1} and \ref{4.2} to the rotating case was proved in  \cite{LeS}:

\begin{thm}[GP limit and BEC at fixed 
$\mathbf\Omega_{\rm rot}$]

In the limit $N\to\infty $ with $g=4\pi Na/L_{\rm trap}$ and $\mathbf\Omega_{\rm rot}$ {\it fixed} the ground state energy of \eqref{84} converges to the minimum energy of the
{\em Gross-Pitaevskii energy functional with rotation}:
\begin{eqnarray}
\label{gpfunc}
\mathcal E^{\rm
GP}[\psi]&=&\int_{\R^3}\left\{\half |\nabla\psi|^2+V|\psi|^2-{\mathbf\Omega_{\rm rot}}\cdot\bar\psi\, {\bf L}\psi+
g|\psi|^4\right\}\hbox{\rm d}^3{\mathbf x}\nonumber \\
&=&\int_{\R^3}\left\{\half |({\rm i}\nabla+\,{\mathbf A})\psi|^2+(V-\hbox{$\frac 12$}\Omega_{\rm rot}^2r)|\psi|^2+
g|\psi|^4\right\}\hbox{\rm d}^3{\mathbf x}
\end{eqnarray}
with the normalization condition
${\int}_{\R^3}|\psi|^2=1$.

Moreover, there is (possibly fragmented) BEC in this limit in the sense that every one-particle density matrix obtained as the limit of normalized one-particle density matrices of ground states of  (\ref{85}), is a convex combination of projectors onto minimizers of the GP functional.
\end{thm}

Every GP minimizer solves the {\em GP equation}
\beq \label{89}\left\{-(\nabla-{\rm i}\,{\mathbf A})^2+(V-\hbox{$\frac 14$}\Omega_{\rm rot}^2r^2)+2
g|\psi|^2\right\} \psi=\mu^{\rm GP} \psi,
\eeq  
but in contrast to the non-rotating case, the minimizer need not be unique up to a constant phase factor. The reason is a  new feature compared to the non-rotating case, namely the  possible occurrence of \emph{vortices} that may break rotational symmetry, even if $V$ depends only on $r$ besides $x_3$.

The proof of the GP limit theorem in \cite{LiS} uses the technique of coherent states and is rather different from the proof in \cite{LSY1,LSe} for the non-rotating case. The reason is that the splitting of space into boxes where the system is approximately homogeneous, as used in the previous proof,  is not applicable in the presence of the global vector potential $\mathbf A$.

The GP equation \eqref{89} and its vortex solutions is a subject of its own that can be studied independently of the many-body problem. See  the monograph \cite{A} and the review article \cite{Fe1} where a large number of references can be found. The most detailed results are for the two-dimensional GP equation, i.e., when $\psi$ depends only on the coordinates $(x_1,x_2)$ in the plane perpendicular to the angular velocity, and in the asymptotic regime when $g\to\infty$. In particular, the  two-dimensional GP equation with a quadratic trap potential $V(\rv)\sim r^2$ has been studied in \cite{AD, IM1, IM2}. More general homogeneous trapping potentials are discussed, e.g., in \cite{CDY2, RD}. In this regime powerful techniques, in particular from Ginzburg-Landau theory  \cite{SS2} (`vortex ball constructions'), can be applied. The case of a fixed, finite value of $g$ is much less explored but several important  general results were obtained in \cite{Seir}. 

It is convenient and customary to write the coupling strength $g=4\pi Na/L_{\rm trap}$ as $1/\eps^2$ with
\beq \varepsilon=g^{-1/2}\eeq
which  is small if $Na/L_{\rm trap}$ is large. This parameter can be thought of as the ratio of the healing length $\ell_c\sim (a\bar\rho)^{-1/2}$ at mean density $\bar \rho=N/L_{\rm trap}^3$ to the length scale $L_{\rm trap}$ of the trap.\smallskip

In anharmonic traps, where $\Omega_{\rm rot}$ can be arbitrary, we shall in particular be interested in the  asymptotic regime where  both $g$ {\it and}  
$\Omega_{\rm rot}$ are large. 


\subsubsection{Status of GP for rapid rotation}

The rigorous derivation of the GP equation from the many-body problem has so far only been achieved for $\Omega_{\rm rot}$ and $\eps$ {\em fixed\/}. 
For rapid rotation
the GP description can break down both in harmonic and anharmonic traps, because the convergence of the many-body quantities to the GP quantities need not hold uniformly in the parameters. The exact limitations, that  may depend on the quantities of interest, have not yet been established rigorously. For instance, even in the non-rotating case BEC has not been proved in the TF limit, i.e., when $g\to \infty$, although there is a limit theorem for the energy and density (cf. Section 4.2), provided the gas remains dilute in the limit. For rotating gases in anharmonic traps an analogous result was proved in \cite{BCPY}:

\begin{thm}[TF limit with rotation]  If $N\to\infty$ with $\Omega_{\rm rot} \to\infty$ and $\eps\to 0$ but the gas remaining dilute (in the sense that  mean density is $\ll a^{-3}$), then  the TF approximation, i.e., the GP energy functional without the kinetic term $\half |({\rm i}\nabla+\,{\mathbf A})\psi|^2$, gives the leading term in the ground state energy as a function of $\Omega_{\rm rot}$ and $\eps$. \end{thm}

The leading TF term, however, does not exhibit vortices which are due to the kinetic term and have only an effect on the energy to next than leading order.

In harmonic traps the limit $\Omega_{\rm rot}\to\Omega_{\rm trap}$ has so far been studied in two steps: In the first step a limit has been considered in which an effective 2D many-body model in the lowest Landau level with contact interactions emerges \cite{LeS}. In the second step sufficient conditions for the validity of a GP limit for this effective model have been derived \cite{LSY}. This will be discussed in more detail in Section 6.2.  A direct derivation of the GP energy functional in the lowest Landau level from the full 3D many-body Hamiltonian \eqref{84} has not yet been carried out.

\subsection{Two-Dimensional GP Vortices}

This subsection provides some heuristic background for understanding the occurrence of vortices in the case of the two-dimensional GP equation. 

The first thing to note is that for sufficiently  small rotational velocities the condensate stays at \emph{rest in the inertial frame} and thus appears to \emph{rotate opposite to $\mathbf\Omega_{\rm rot}$ in the rotating frame}. This is a manifestation of superfluidity:  A  normal fluid would pick up the rotational velocity of the container and in equilibrium the fluid would be  at rest in the rotating frame.

In the rotating frame the operator of the  velocity is {$-{\rm i}\nabla-{\bf A}(\bf r)$}. The constant wave function, that (in a `flat' trap) minimizes the GP energy functional (with zero energy in excess of the interaction energy) for small $\Omega_{\rm rot}$, thus has, in the rotating frame,  the velocity
\beq {\mathbf v}(\mathbf r)=-{\bf A({\mathbf r})}=-{\mathbf \Omega_{\rm rot}}\times{\mathbf r}=-\Omega_{\rm rot} r \,{ {\mathbf e}_\theta},\eeq  
where ${\mathbf e}_\theta$ denotes the unit vector with respect to the angular variable.
Note that the kinetic energy corresponding to this velocity is exactly compensated by the centrifugal term $-\hbox{$\frac 12$}\Omega_{\rm rot}^2r^2$ in the GP energy functional (\ref{gpfunc}). 

At higher rotational velocities the condensate responds by creating vortices whose velocity field may partly compensate the term $-\bf A$ of the velocity and hence reduce the kinetic energy. This reduction does not come for free, however, because the creation of a vortex is accompanied by a redistribution of the density and hence an increase in interaction energy.

To estimate these competing effects let us consider the case of  large $g$, i.e., small  $\varepsilon=g^{-1/2}$ and a trap with effective radius $R$.  A  \emph{vortex of degree $d$}  located at the origin can, for the purpose of this heuristic discussion,  be approximated by the ansatz
\beq \psi(r,\theta)=f(r)\exp({\rm i}\theta d)\eeq   
with 
\beq
        	f(r) \sim
        	\left\{
        	\begin{array}{ll}
            		r^d    &   \mbox{\rm if} \:\:\:\: 0\leq r\lesssim r_{\rm v}  \\
            		\mbox{}     &   \mbox{} \\
            		R^{-1}     &   \mbox{\rm if} \:\:\:\: r_{\rm v} \lesssim r\leq R       	\end{array}
        	\right.
    	\eeq
	where $r_{\rm v}$ is the radius of the vortex core where the density is small. Now the component of the velocity in the direction of ${\mathbf  e}_\theta$ is
	\beq \mathbf v(\mathbf r)_\theta=\left(\frac dr-\Omega_{\rm rot} r\right){\mathbf  e}_\theta\ .\eeq  
	The change in kinetic energy compared to the vortex free case, $d=0$, is therefore
\begin{equation}\sim R^{-2}\int_{r_{\rm v}}^R[(d/r)^2-d\,\Omega_{\rm rot}]\,r\,dr+O(1)\\=R^{-2}d^2|\log(r_{\rm v}/R) |-\half d\, \Omega_{\rm rot}+O(1).\end{equation}
On the either hand the change in interaction energy through the creation of the vortex is
\beq \sim\frac 1{\eps^2} (r_{\rm v}/R)^2.\eeq
Optimizing the total energy change w.r.t. $r_{\rm v}$ gives a  vortex radius of the order of the healing length, i.e.
\beq r_{\rm v}\sim\eps R\eeq
and an interaction energy increase $\sim R^{-2}$. A vortex of degree $d=1$ becomes energetically favorable when this is outweighed by a decrease in kinetic energy, i.e., if
\beq R^{-2}|\log \varepsilon|-\half \Omega_{\rm rot}+O(1)R^{-2}<0\eeq  
which means
\beq \Omega_{\rm rot}\gtrsim O(1)R^{-2}|\log\varepsilon|.\eeq  
We also see that $d$ vortices of degree 1, ignoring their interaction, have energy $\sim d(R^{-2}|\log \varepsilon|- \Omega_{\rm rot})$ while a vortex of degree $d$ has energy $R^{-2}d^2|\log\varepsilon |- d\,\Omega_{\rm rot}$. Hence it is energetically favorable to \lq split' a $d$-vortex into $d$ pieces of 1-vortices, breaking the rotational symmetry. 
    
             These heuristic considerations are confirmed by a detailed analysis for \lq slowly' rotating gases, i.e., 
          $\Omega_{\rm rot}=O(|\log \varepsilon|)$ \cite{AD, AJR, IM1, IM2, RD}.
In a `flat' trap vortices start to appear for {$\Omega_{\rm rot} R^2=\pi |\log \varepsilon |$} and for
	\beq\label{vortexcond} |\log \varepsilon |+(d-1)\log|\log\varepsilon|<\Omega_{\rm rot} R^2/\pi\leq  |\log \varepsilon |+d\log|\log\varepsilon|\eeq there are exactly $d$ vortices of degree 1. 
In a homogeneous trap, $V(r)\sim r^s$, the effective radius $R$ of the condensate can be estimated by equating the potential energy in the trap and the interaction energy, i.e., 
	$R^s\sim \varepsilon^{-2} R^2(R^{-1})^4$, which leads to 
	 \beq R\sim \varepsilon^{-2/(s+2)}\eeq   and thus the critical velocity for the creation of a vortex is
	\beq \sim \varepsilon^{4/(s+2)}|\log\varepsilon|.\eeq  
	In particular, for a harmonic trap with $s=2$ the critical rotational velocity  is $\sim \eps|\log\eps|$.
	The decrease with $\eps$ of the critical velocity for creating a vortex in such traps is entirely due to the fact that the interaction spreads out the mass and decreases the density. In a trap with hard walls so that the radius is fixed the critical velocity is $\sim|\log\eps|$ and hence increases when $\eps\to 0$.


\subsection{GP Theory for Rapid Rotation, Anharmonic Traps}

In this subsection we consider the effects of rapid rotation on the GP minimizer in a 2D anharmonic trap. It should be noted that 2D results are of experimental relevance also in 3D: 1) to systems that are strongly confined in one direction so that the motion is effectively two-dimensional, or 2) for traps that are highly elongated in the direction of the rotational axis so that the properties of the condensate are approximately independent of the coordinate in this direction. \smallskip

For mathematical simplicity we consider a  2D \lq flat', disc-shaped trap  with  rigid boundary and radius 1. Some comments on more general traps will be made at the end.
  
The GP energy functional on the unit disc $\mathcal D\subset \mathbb R^2$ is
\beq \mathcal E^{\rm GP}[\psi]=\int_{\mathcal D}\left\{\half |({\rm i}\nabla+\,{\mathbf A})\psi|^2-\hbox{$\frac 12$}\Omega_{\rm rot}^2r^2|\psi|^2+
\frac1{\eps^2}|\psi|^4\right\}\mathrm d^2\rv\eeq
where  ${\mathbf A}(\rv)= \Omega_{\rm rot}\,r\,{\mathbf e}_\theta$. \smallskip
 
As mentioned in the previous subsection one can prove that if $\Omega_{\rm rot}\leq \Omega_{c_1}|\log\eps| +O(\log|\log\eps|)$ for a certain $\Omega_{c_1}$ there is a  finite number of vortices, even as $\eps\to 0$. For larger $\Omega_{\rm rot}$ the number of vortices is  unbounded as $\eps\to 0$. If $\Omega_{\rm rot}$ is still $O(|\log \eps|)$ the vortices are not uniformly distributed, however.  This transition region has recently been analyzed in \cite{CR}.\smallskip

For $\Omega_{\rm rot}\gg |\log\eps|$ new phenomena appear at two critical velocities, namely for $\Omega_{\rm rot}\sim 1/\eps$ and $\Omega_{\rm rot}\sim 1/(\eps^2|\log\eps|)$ respectively:

If $\Omega_{\rm rot}=O(1/\eps)$ the centrifugal term $-(\Omega_{\rm rot}^2/2) r^2|\psi(\rv)|^2$ and the interaction term $(1/\eps^2)|\psi(\rv)|^4$
 are comparable in size and the centrifugal forces influence the bulk shape of the condensate. The kinetic energy term $\half |(\mathrm i\nabla +{\mathbf A}(\rv))\psi(\rv)|^2$ is formally also of order $1/\eps^2$ if 
  $\Omega_{\rm rot}\sim 1/\eps$, but it turns out that its contribution to the energy is, in fact, of  lower order, namely $\sim\Omega_{\rm rot}|\log\eps|$, because  a lattice of vortices  emerges as $\eps\to 0$. The velocity field generated by the vortices compensates partly the field $-{\mathbf A}(\rv)$ generated by the rotation.
  
  For $\Omega_{\rm rot}\geq \Omega_{c_2} 1/\eps$  the centrifugal forces deplete strongly the density  in a `hole' of radius 
\beq R_{\rm h}=1-c (\Omega_{\rm rot}\eps)^{-1}\eeq
  around the rotation axis and the bulk of the condensate is concentrated in a thin annulus of thickness $\sim(\eps\Omega_{\rm rot})^{-1}$.  As long as 
  $\Omega_{\rm rot}\ll 1/(\eps^2|\log\eps|)$, however,  the annulus still contains a lattice of vortices, but if $\Omega_{\rm rot}>\Omega_{c_3}1/(\eps^2|\log\eps|)$ the high density of the condensate in the annulus make vortices too costly. A transition to a {\it `giant vortex'} state \cite{KTU, FB, KB, FJS, FZ} takes place where all vorticity is concentrated in the `hole' but the bulk of the condensate is vortex free.\smallskip

We now discuss these results in more detail, starting with the parameter region  $|\log\eps|\ll \Omega_{\rm rot}\ll 1/(\eps^2|\log\eps|)$ where the following holds \cite{CY}:
 
\begin{thm}[Energy to subleading order] \label{5.3}
 Let $E^{\rm GP}$ denote the GP energy, i.e., the minimum of the GP energy functional. Let $E^{\rm TF}$ denote the minimal energy of the GP functional {\em  without} the kinetic term. \smallskip
  
 If $|\log\eps|\ll \Omega_{\rm rot}\ll 1/\eps$, then
 \beq\label{105} E^{\rm GP}=E^{\rm TF}+\half\Omega_{\rm rot}|\log(\eps^2\Omega_{\rm rot})|(1+o(1)).\eeq
 
  If $1/\eps\lesssim  \Omega_{\rm rot}\ll 1/({\eps^2}|\log\eps|)$  then
  \beq\label{106} E^{\rm GP}=E^{\rm TF}+\half \Omega_{\rm rot}|\log\eps|(1+o(1)).\eeq
  In both cases the energy corresponds to a uniform distribution of vorticity in a the form of a vortex lattice in the bulk of the condensate.\end{thm}


\subsubsection{An electrostatic analogy}

 The upper bound to the energy in Theorem 5.3 is based on a variational ansatz that is motivated by an electrostatic analogy. \smallskip
 
 We write points $\rv=(x,y)\in \mathbb R^2$ as complex numbers, $\zeta=x+{\mathrm i}y$, and consider a lattice of points $\zeta_j$. Placing a vortex of degree 1 at each point $\zeta_j$ leads to a trial function for the GP energy of the form
  \beq \psi(\rv)=f(\rv)\exp\{{\rm i}\varphi(\rv)\}\eeq
  where $f$ is real valued with a zero at each of the points $\zeta_j$ and the phase factor is
  \beq\exp\{{\rm i}\varphi(\rv)\}=\prod_j\frac{\zeta-\zeta_j}{|\zeta-\zeta_j|}.\eeq
Now
\beq |(\mathrm i\nabla+\mathbf A)\psi|^2=|\nabla f|^2+f^2|\mathbf A-\nabla\varphi|^2\eeq
and 
\beq \varphi=\sum_j\arg (\zeta-\zeta_j).\eeq 
The phase $\arg z$ of a complex number is the imaginary part of the complex logarithm which is an analytic function on the complex plane (suitably cut). The Cauchy-Riemann equations for the real and imaginary part of an analytic functions imply
 \beq |\mathbf A-\nabla\varphi|^2=|\Omega_{\rm rot} \, r\mathbf e_r-\nabla \chi|^2\eeq
 where
 \beq \chi(\rv)=\sum_j\log|\rv-\rv_j|.\eeq

 But
  \beq {\mathbf E}(\rv):= \Omega_{\rm rot}\,r{\mathbf e}_r-\nabla\chi(\rv)\eeq
has a simple physical interpretation: It can be regarded as an \lq electric field' generated by a  uniform charge distribution of density $\Omega_{\rm rot}/\pi$ together with  unit \lq charges' of opposite sign at the positions of the vortices, $\rv_j$. The integral of $|\mathbf E(\rv)|^2$ is the corresponding electrostatic energy.\smallskip


To construct a trial function for an upper bound to the GP energy 
we  distribute the  vortices over the unit disk so that the  vorticity per unit area is  $\Omega_{\rm rot}/\pi$. (This is really $2\Omega_{\rm rot}\cdot m/h$.) Thus every vortex sits at the center $\rv_j$ of  lattice cell $Q_j$ of area $|Q_j|= \pi/\Omega_{\rm rot}$, surrounded by a uniform charge distribution of the opposite sign so that the total charge in the cell is zero.\medskip
 
If the cells were disc-shaped, then Newton's theorem would imply that the 'electric field' generated by each cell would vanish outside the cell, i.e, there would be no interaction between the cells. 
The cells are, of course, never strictly disc shaped, but among the three possibilities (triangular, rectangular and hexagonal)  for tilings of the plane with regular polygonal cells the {\em hexagonal} ones have the least mutual interaction energy. The vortices then sit on a {\it triangular lattice}. The interaction between the cells, although not zero, is small because the cells have only a quadrupole moment or higher and  no dipole moment. The hexagonal cell is distinguished by having the smallest multipole moments.\footnote{Although this is in accord with the apparent arrangement of vortices as observed in experiments, it is not a proof that the vortices {\it must} sit on a triangular lattice. First, it is not proved that the vortices sit on a regular lattice at all, and secondly, it turns out that the order of the energy considered in Theorem \ref{5.3} does not distinguish between different regular lattices that all lead to cells with vanishing dipole moment. The energetic distinction between different regular arrangements is a delicate higher order effect.  This topic has recently been analyzed in \cite{SS}.}

The upper bound to the energy proved in \cite{CY} was achieved along the line just described. The lower bound relies on constructions and theorems from Ginzburg-Landau theory \cite{SS2}.


\subsubsection{Emergence of a \lq giant vortex'}

As already mentioned,  a transition to a new phase takes place when $\Omega_{\rm rot}$ is of the order $1/(\eps|\log\eps|^2)$. Here a variational ansatz of the form
\beq\label{114} \psi(\mathbf r)=f(\mathbf r)\exp(\mathrm i\hat\Omega_{\rm rot}\theta)\eeq
 with a real valued function $f$ and
\beq \hat\Omega_{\rm rot}=\Omega_{\rm rot}-O(\eps^{-1})\eeq
gives a lower energy that the vortex lattice ansatz leading to \eqref{105}, \eqref{106}. This does not prove, however, that the energy $E^{\rm gv}$ of the ansatz \eqref{114} gives a good approximation to the energy of the true minimizer, nor that the latter is free of vortices in the bulk. That both statements are true is the content of the following theorems \cite{CRY}: 

\begin{thm}[Energy in the giant vortex regime] There is a constant  $0<\Omega_{\rm c_3}<\infty$ such that for $\Omega=\Omega_0(\eps|\log\eps|^2)^{-1}$ with $\Omega_0>\Omega_{c_3}$ the ground state energy is
 \beq E^{\rm GP}\label{gvenergyas}=E^{\rm gv}-O(|\log\eps|^{3/2}/\eps^2).\eeq
 \end{thm}
\begin{thm}[Absence of vortices in the bulk]\label{5.5}
There is an annulus $\mathcal A$ of width $O((\eps\Omega_{\rm rot})^{-1})$ with $\int_{\mathcal A}|\psi^{\rm GP}|^2=1-o(1)$ such that for
$\Omega$ as above and $\eps$ sufficiently small 
the minimizer 
$\psi^{\rm GP}$ is free of zeros in the annulus.
\end{thm}
 
The proof, in particular of Theorem \ref{5.5},   is surprisingly difficult
but a heuristic explanation for the transition at $\Omega_{\rm rot}\sim1/(\eps^2|\log\eps|)$ can be given by exploiting the electrostatic analogy:
 
 Consider the variational ansatz \eqref{114} and interpret $\hat\Omega_{\rm rot}$ as a `charge' situated at the origin. 
The `electric field' generated this charge exactly cancels, in the annulus $\mathcal A$, the `electric field'
generated in the annulus by the uniform charge density $\Omega_{\rm rot}/\pi$ of the `hole' (by Newton's theorem), due to the vector potential. However, the `charge' corresponding to the vector potential  in the annulus is not cancelled, and this 'residual charge' is
\beq \sim \Omega_{\rm rot} \times (\eps\Omega_{\rm rot})^{-1}=\eps^{-1}.\eeq
The electrostatic energy of this residual charge distribution is $\sim \eps^{-2}$.

Creating a vortex in the annulus neutralizes one charge unit and thus reduces the electrostatic energy by $\eps^{-1}$. 
On the other hand, the {\em cost} of a vortex is $\sim f^2\,|\log\eps|$, and we have $f^2\sim(\eps\Omega_{\rm rot})$, so the cost of a single vortex in the bulk is
\beq \sim\eps\Omega_{\rm rot}\,|\log\eps|\,.\eeq
Gain and cost are comparable if $\eps^{-1}\sim \eps\Omega_{\rm rot}\,|\log\eps|$, i.e., for
\beq \Omega_{\rm rot}\sim \frac{1}{\eps^2|\log\eps|}.\eeq
 If $\Omega_{\rm rot}$ is smaller it still pays to create vortices also in the annulus, but if $\Omega_{\rm rot}$ is larger, the cost outweighs the gain and the annulus is vortex free.

This picture is substantiated  by the rigorous analysis in \cite{CRY} and it has been generalized to other anharmonic traps besides the `flat' case in \cite{CPRY2}. Here some interesting new features concerning the size of the vortices and the shape of the density distribution in the annulus appear. See  \cite{Y, CPRY3} for surveys on this topic. 

\begin{figure}[htf]
\center
\fbox{\includegraphics[width=10cm]{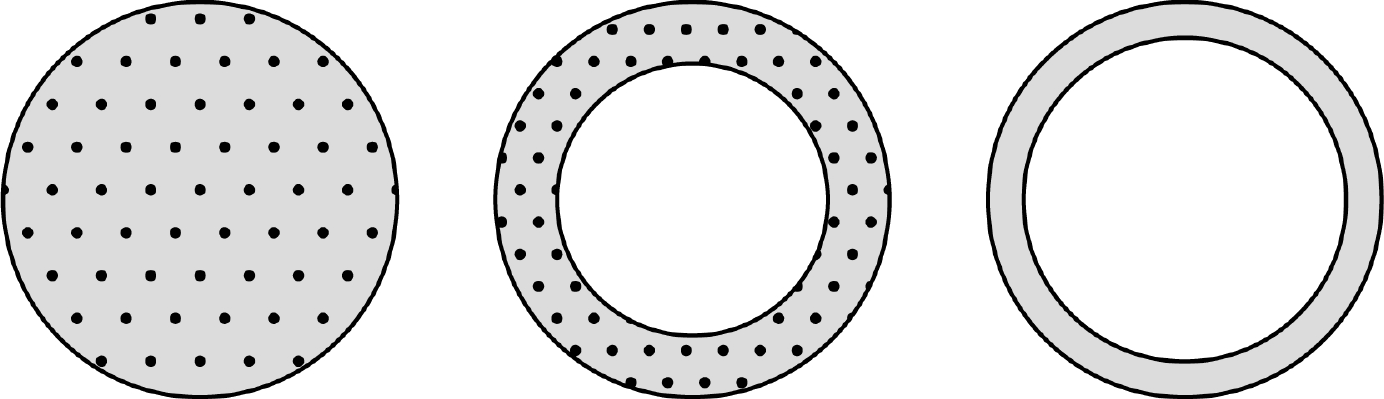}}
\caption{The transition to a giant vortex.}
\label{fig3}
\end{figure}

\subsubsection {Summary on vortices}

The emergence of single vortices, vortex lattices and a `giant vortex' state for a condensate in a rapidly rotating
anharmonic trap can be understood by asymptotic analysis of the GP equation. When both the coupling constant $1/\eps^2$ and the rotational velocity $\Omega_{\rm rot}$ are large the picture is as follows (in a `flat' trap):

\begin{itemize}
\item Single vortices for $\Omega_{\rm rot}\sim| \log \eps|$\medskip
\item A vortex lattice for $|\log\eps|\ll \Omega_{\rm rot} \ll 1/(\eps^2|\log\eps|)$\medskip
\item A `hole' due to centrifugal forces for $1/\eps\lesssim \Omega_{\rm rot}$.\medskip
\item A `giant vortex' for $1/(\eps^2|\log\eps|)\lesssim \Omega_{\rm rot}$
\end{itemize}

\section{Rapid Rotation and Confinement to the Lowest Landau Level}


 One of the most striking phenomena in condensed matter physics is the Fractional Quantum Hall Effect (FQHE) for charged fermions in strong magnetic fields \cite{STG} that still, after decades of research, poses many challenging questions. It has been recognized for some time that {\em bosonic} analogues of the FQHE can be studied in cold quantum gases set in rapid rotation, see \cite{Vie, Co} and references cited therein.
In this section the focus will be on {\it one} aspect of the Quantum Hall Physics of cold bosons: The emergence of {\it strongly correlated many-body states} through appropriate tuning of the parameters. 
 
 The starting point is the many-body Hamiltonian \eqref{84} in the rotating frame that we recall for convenience:
 \begin{equation}
H_N= 
\sum_{j=1}^{N} \left(- \half \nabla^2_j +V(\x_{j})-{\mathbf L}_j\cdot {{\mathbf \Omega}}\right)+
\sum_{1\leq i < j\leq N } v(|\x_i - \x_j|).
\end{equation}
  
 In contrast to Section 5 we now consider the case when $V$ is a quadratic potential in the direction $\perp$ to the rotation axis,
 \beq V(\x)=\half \Omega_{\rm trap}^2 r^2+V^{\parallel}(x_3),\eeq
 with $r^2=x_1^2+x_2^2$ and the angular velocity $\Omega_{\rm rot}$ approaches the frequency of the potential  $\Omega_{\rm trap}$ from below. 
When writing the  Hamiltonian with this potential in a `magnetic'  form, it is convenient to define the vector potential in a different way than in \eqref{vecpot}, namely
 \beq \mathbf A(\x)=\Omega_{\rm trap}(x_2,-x_1,0)\eeq
 rather than $\mathbf A(\x)=\Omega_{\rm rot}(-x_2,x_1)$ as in \eqref{vecpot}. With this definition the Hamiltonian takes the form
 \beq H_N=\sum_{j=1}^N\left\{\half (\mathrm i\nabla_j+\mathbf A(\x_j))^2+\omega\, \mathbf e_3\cdot \mathbf L_j+V^{\parallel}(x_3)\right\}+  \sum_{ i < j } v(|\x_i - \x_j|) \eeq
with
\beq \omega:=\Omega_{\rm trap}-\Omega_{\rm rot}>0.\eeq


\subsection{Confinement to the Lowest Landau Level, 1-Particle Case}

Consider now the one-particle Hamiltonian
\beq H_1=\half (\mathrm i\nabla_\perp+\mathbf A(\x))^2+\omega \mathcal L-\half\partial_3^2+ V^\parallel(x_3)\eeq
where we have written $\mathcal L=\mathbf e_3\cdot \mathbf L$ and $\nabla_\perp=(\partial_1,\partial_2)$.
This is a sum of three commuting operators,
\beq \half (\mathrm i\nabla_\perp+\mathbf A(\x))^2,\quad \omega \mathcal L\quad\text{and}\quad -\half\partial_3^2+ V^\parallel(x_3).\eeq
 The spectrum of $\half (\mathrm i\nabla_\perp+\mathbf A(\x))^2$ is
\beq (n+\half)\mathrm 2\Omega_{\rm trap},\quad n=\mathrm{0,1,2},\dots, \quad \mathrm{called}\ \ \text{{\it Landau levels.}}\eeq
The spectrum of $\omega \mathcal L$ is
\beq \ell\omega, \quad \ell=0,\pm 1,\pm 2\dots\eeq
 and $-\half\partial_3^2+ V^\parallel(x_3)=: h^\parallel$  has a spectral gap, $e^\parallel>0$, above its ground state.\smallskip

For $\omega\ll\min\{\Omega_{\rm trap}, e^\parallel \}$  it is natural to restrict attention to states with $n=0$, and the motion in the $x_3$-direction is `frozen' in the ground state of $h^\parallel$. 
From now on we choose units so that  $\Omega_{\rm trap}=1$.\footnote{Since $|\nabla\times \mathbf A|=2\Omega_{\rm trap}$ this means that the `magnetic field' unit is 2 rather than 1. For this reason some formulas in the sequel may differ by powers of 2 from the ones customary in the theory of the FQHE for electrons.}\smallskip

   Replacing $(x_1,x_2)$ by the complex coordinate $z=x_1+\mathrm i x_2$   and denoting
$\partial=\half(\partial_{1}-\mathrm i\partial_{2})$, $\bar \partial=\half(\partial_{1}+\mathrm i\partial_{2})$
we can write
\beq \half (\mathrm i\nabla_\perp+\mathbf A(\x))^2=2\left(\hat a^\dagger \hat a+\half\right)\eeq
with 
\beq \hat a^\dagger:=\half(-2\partial+\bar z),\qquad \hat a:=\half (2\bar \partial+ z).\eeq
These operators satisfy the canonical commutation relations \beq [\hat a,\hat a^\dagger]=1.\eeq
The operators 
\beq \hat b^\dagger:=\half (-2\bar \partial+ z),\qquad \hat b:=\half (2\partial+ \bar z) \eeq 
also satisfy the canonical commutation relations and commute with $\hat a$ and $\hat a^\dagger$. They correspond to a replacement $\mathbf A\to-\mathbf A$:
\beq \half (\mathrm i\nabla_\perp-\mathbf A(\x))^2=2(\hat b^\dagger \hat b+\half).\eeq
Moreover,
\beq \hat b^\dagger \hat b-\hat a^\dagger \hat a=z\partial-\bar z\bar \partial =\mathcal L.\eeq
Hence the eigenvalues of $\hat a^\dagger \hat a$ are infinitely degenerate and  the degenerate eigenstates can be labelled by  eigenvalues of  either  $\hat b^\dagger \hat b$ or  $\mathcal L$.

\subsubsection{Bargmann space}

 The lowest eigenvalue of $\hat a^\dagger\hat  a$ is zero. The corresponding eigenfunctions $\psi(z,\bar z)$ are solutions of  the equation $\hat a\psi=0$, i.e.,
   \beq \bar \partial \psi(z,\bar z)=-\half z\psi(z,\bar z).\eeq
   Thus
   \beq\psi(z,\bar z)=\varphi(z)\exp(-|z|^2/2)\eeq
with $\bar \partial \varphi(z)=0$, i.e., $\varphi$ is an {\it analytic} function of $z$.
   
 In the lowest Landau level (LLL) we are thus led to consider the {\it Bargmann space} $\mathcal B$ \cite{Bar, GJ} of  analytic functions $\varphi$ such that
 \beq \langle \varphi,\varphi\rangle:=\int |{\varphi(z)}|^2 \exp(-|z|^2)\,\mathrm d^2z<\infty\eeq
   where $\,\mathrm d^2z$ denotes the Lebesgue measure on $\mathbb C$ (regarded as $\mathbb R^2$).

 The Bargmann space is a Hilbert space with scalar product
    \beq \langle \varphi,\psi\rangle:=\int \bar \varphi(z) \psi(z)\exp(-|z|^2)\,\mathrm d^2z.\eeq
    On $\mathcal B$ the angular momentum operator is  $\mathcal L=z\partial$.    
    Moreover, for $\varphi\in\mathcal B$,
    \beq \langle\varphi,\mathcal L\varphi\rangle=\int ( |z|^2-1)|{\varphi(z)}|^2 \exp(-|z|^2)\,\mathrm d^2z .\eeq
    The eigenvalues of $\mathcal L$ restricted to $\mathcal B$ are $\ell=0,1,2,\dots$ with corresponding normalized eigenfunctions 
   \beq\varphi_\ell(z)=\left(\pi \ell!\right)^{-1/2}\, z^\ell.\eeq
  Note that $|\varphi_\ell(z)|^2 e^{-|z|^2}$ has in the radial variable a maximum  at $r_\ell=\sqrt{\ell}$. The density of states per unit area is therefore $1/\pi$.

\subsection{Confinement to the Lowest Landau Level, N-particle Case}

We now come to the $N$-body problem for bosons in the LLL. The relevant Hilbert space is 
\beq \mathcal B_N=\mathcal B^{\otimes^N_{\rm symm}},\eeq
 i.e., it consists of symmetric, analytic functions $\psi$ of $z_1,\dots,z_N$ such that
 \beq\label{144} \int_{\mathbb C^N}|\psi(z_1,\dots,z_N)|^2\exp\Big(-\sum_{j=1}^N|z_j|^2\Big)\,\mathrm d^2z_1\cdots \,\mathrm d^2z_N<\infty.\eeq 
 
  As next we take the {\it interaction} into account, i.e., consider a suitable image of 
  $\sum_{1 \leq i < j \leq N} v(|\x_i - \x_j|)$ as an operator on $\mathcal B_N$.


\subsubsection{Contact interaction}

 For short range, nonnegative interaction potentials $v$ it was shown in \cite{LeS}  that  for  $\omega a\ll1$, with $a$ the scattering length of  $v$ the motion is indeed restricted to the 2D  LLL, and moreover that  $v(\x_i-\x_j)$  can be replaced by $g\,\delta(z_i-z_j)$ with\footnote{Note that this $g$ is, by definition, not proportional to $N$, in contrast to  the previous coupling constant $g$ in \eqref{48}.}\beq g\sim a\sqrt{e^\parallel}>0.\eeq  
 Such a contact potential is perfectly acceptable for analytic functions and 
is even given by a bounded operator on the Bargmann space:
 
  Define $\delta_{12}$ on $\mathcal B_2$ by
   \beq\delta_{12}\varphi(z_1,z_2)=\frac 1{2\pi}\varphi\big(\half(z_1+z_2), \half(z_1+z_2)\big).\eeq
Then a simple computation, using the analyticity of $\varphi$, shows that
\beq \langle \varphi,\delta_{12}\varphi\rangle=\int_{\mathbb C}|{\varphi(z,z)}|^2\exp(-2|z|^2)\,\mathrm d^2z.\eeq

Replacing $v$ by the contact interaction, the effective Hamiltonian on Bargmann space  becomes (apart from an additive constant)
\beq\label{yrastham} H_N^{\rm 2D}= {\omega}\,\mathcal L_N+{g}\,\mathcal I_N\eeq 
with
\beq \mathcal L_N=\sum_{i=1}^Nz_i\partial_i\, \quad\quad \mathcal I_N=\sum_{i<j}\delta_{ij}.\eeq
An important feature of the Hamiltonian \eqref{yrastham}  is that the operators $\mathcal L_N$ and $\mathcal I_N$ commute. The  lower boundary of (the convex hull of) their joint spectrum in a plot with angular momentum as the horizontal axis is called the  {\it yrast curve}. See Fig. 4 and  \cite{Vie} for its qualitative features. 

As a function of the eigenvalues $L$ of $\mathcal L_N$ the yrast curve $I(L)$ is decreasing from $I(0)= (4\pi)^{-1} N(N-1)$ to $I(N(N-1))=0$. The monotonicity follows from the observation that if a simultaneous eigenfunction of $\mathcal L_N$ and $\mathcal I_N$ is multiplied by the center of mass, $(z_1+\cdots+z_N)/N$, the interaction is unchanged while the angular momentum increases by one unit. 

For a given ratio $\omega/g$ the ground state of \eqref{yrastham} (in general not unique) is determined by the point(s) on the yrast curve  where a supporting line has slope $-\omega/g$. The ground state energy is
\beq E(N,\omega,g)=\min_L(\omega L+g\,I(L)).\eeq

The {\em filling factor} of a state with angular momentum $L$ is defined as
\beq \nu=\frac{N(N-1)}{2L}=\frac N{N_{\rm v}}\eeq
where $N_{\rm v}=2L/(N-1)$ is the number of vortices.  
The filling factor of the ground state depends on the ratio $\omega/g$ and varies from $\infty$ to 0 as the ratio decreases and the angular momentum increases.


\subsubsection{The GP regime}

A rough estimate for the radius $R$ of the system, assuming that kinetic and interaction energy are of the same order of magnitude, gives \beq R\sim (Ng/\omega)^{1/4}\qquad \hbox{\rm and}\qquad 
L\sim N\Omega R^2\sim N(Ng/\omega)^{1/2}.\eeq
 Thus, if If $\omega/g \gg N^{-1}$ then $L\ll N^2$ and $\nu\gg 1$. 
In this case the ground state can be shown to be well described by an uncorrelated Hartree state
  $\left(\varphi^{\rm GP}\right)^{\otimes N}$ where $N^{1/2}\varphi^{\rm GP}$ minimizes the GP energy functional
\beq\mathcal E^{\rm GP}[\varphi]=\omega\langle\varphi, \mathcal L\varphi\rangle+\frac g2\int_{\mathbb C}|\varphi(z)|^4\exp(-2|z|^2)\,\mathrm d^2z\eeq
under the condition $\int|\varphi|^2\exp(-|z|^2)\mathrm d^2z=N$ with energy \beq E^{\rm GP}(N,\omega, g)=NE^{\rm GP}(1,\omega, Ng).\eeq 
Note that only  analytic functions in the Bargmann space $\mathcal B$ are allowed as trial functions for an upper bound to this energy.

More precisely, the following holds \cite{LSY}:
 
\begin{thm}[GP limit theorem in LLL]  
    For every $c>0$ there is a $C<\infty$ such that 
  \beq E^{\rm GP}(N,\omega,g)\geq E(N,\omega,g)\geq E^{\rm GP}(N,\omega, g)(1-C(g/N\omega)^{1/10})\eeq
  provided $gN/\omega>c$.\end{thm}
  
    The lower bound covers the whole regime $L\ll N^2$, i.e., $\nu\gg 1$, but the GP description might have a wider range of applicability.  
  The proof uses similar techniques as in \cite{LSe} for the 3D GP limit theorem at fixed $\Omega$ and $Na$, in particular coherent states.


\subsubsection{Gaps}

For every value of the angular momentum $L$ the interaction operator $\mathcal I_N$ has a nonzero spectral gap above 0
\beq \Delta_N(L)=\inf \{{\rm spec}\ \mathcal I_N\upharpoonright_{\mathcal L_N=L}\setminus \{0\}\}>0.\eeq
The gap, and hence the Yrast curve $I(L)$, are  monotonously decreasing with $L$ for the reason already mentioned: The angular momentum of an eigenstate of $\mathcal I_N$ can be increased by one unit by multiplying the wave function with the center of mass coordinate $(z_1+\cdots +z_N)/N$. 
This does not change the interaction energy and leads to a family of `daughter states' for each state on the yeast curve.
There is numerical and some theoretical evidence that
\beq\label{157} \Delta_N(L)\geq \Delta_N(N(N-1)-N)=\Delta:=\min_{L'}\Delta_N(L') >0\eeq
for all $L$ independently of $N$ but this is still not proved. We shall call the validity of \eqref{157} the {\it gap conjecture}.

\begin{figure}[htf]
\center
\fbox{\includegraphics[width=12cm, bb=100 450 500 730]{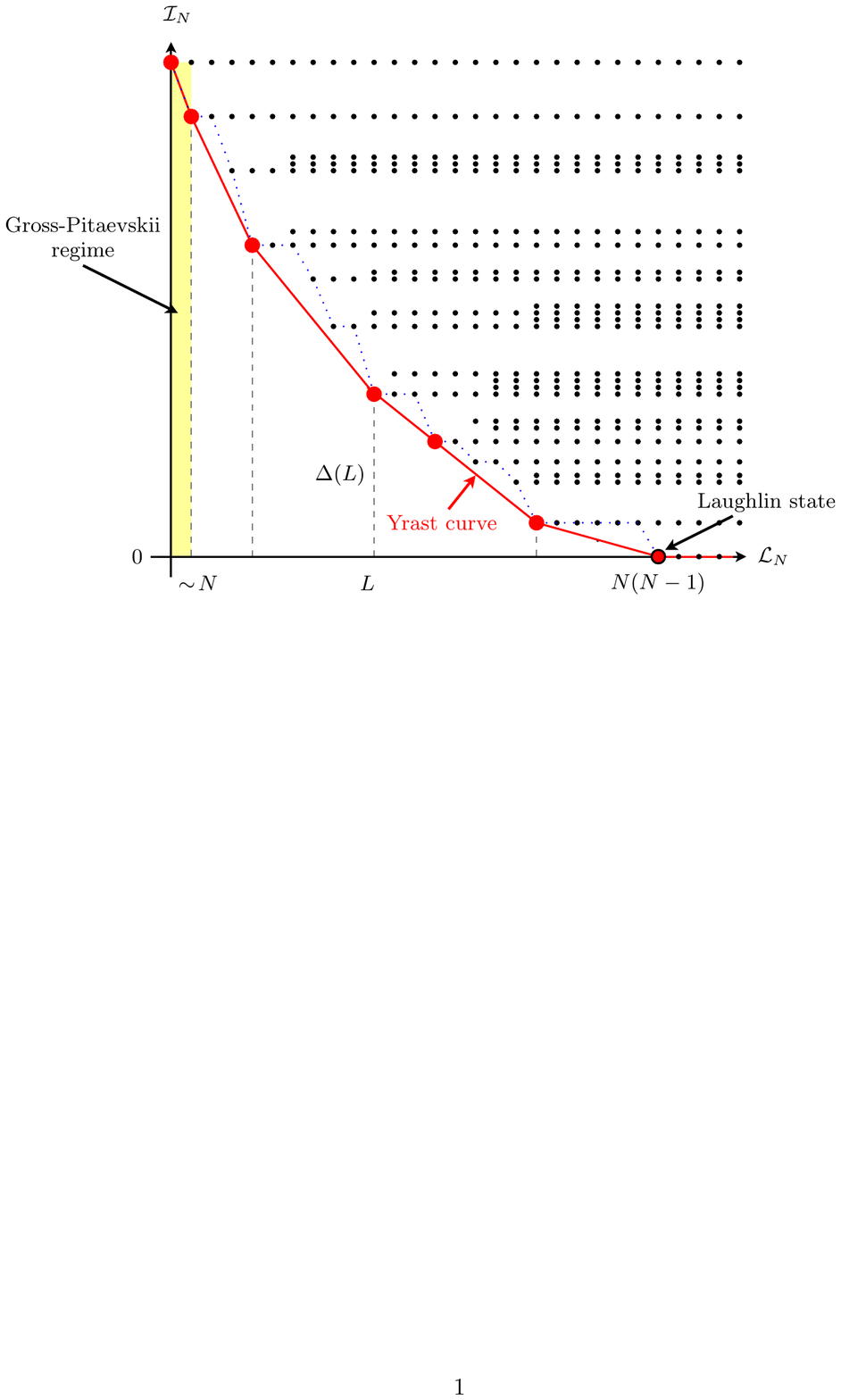}}
\caption{The joint spectrum of $\mathcal L_N$ and $\mathcal I_N$ (adapted from \cite{LeS}).}
\label{fig4}
\end{figure}


\subsection{Passage to the Laughlin state}

As the filling factor decreases the ground state becomes increasingly correlated. The exact ground states are largely unknown (except for $L\leq N$ \cite{PB}), but candidates of states with various rational filling factors ({\it composite fermion states, Moore-Read states, Read-Rezayi states,...}) for energy upper bounds have been suggested and studied, See e.g. the review article \cite{Co}. 

If $\omega/g<\Delta/N^2$ one reaches the {\em Laughlin state} with filling factor $\half$ whose wave function in Bargmann space is
\beq \psi_{\rm Laughlin}(z_1,\dots,z_N)=c\prod_{i<j}(z_i-z_j)^2.\eeq
It has interaction energy 0, angular momentum $L=N(N-1)$.  \smallskip
 
The limit $\omega\to 0$, keeping $\omega>0$, is experimentally very delicate, however.
 For  stability,  but also to study new effects, we consider now a modification of the Hamiltonian by adding a small anharmonic term:
\beq H^{\rm 2D}_N\rightarrow H^{\rm 2D}_N+k\sum_{i=1}^N |z_i|^4\eeq
with a new parameter $k>0$.  The potential $|z|^4$  can be expressed through $\mathcal L$ and $\mathcal L^2$ on Bargmann space, because
with $\mathcal L=z\partial$ we have by partial integration, using the analyticity of $\varphi$,
\beq \langle\varphi,\mathcal L\varphi\rangle=\int |{\varphi(z)}|^2 (|z|^2-1)\exp(-|z|^2)\,\mathrm d^2z \eeq 
 and
\beq \langle\varphi,\mathcal L^2\varphi\rangle=\int (|z|^4-3|z|^2+1)|{\varphi(z)}|^2 
 \exp(-|z|^2)\,\mathrm d^2z. \eeq 
Thus the modified Hamiltonian, denoted again by $H^{\rm 2D}_N$,  can be written (up to an additive constant)
\beq \label{162}H^{\rm 2D}_N=(\omega+{3k})\mathcal L_N+{k} \sum_{i=1}^N \mathcal L_{(i)}^2+g\,\mathcal I_N.\eeq
 
\subsubsection{Fully correlated states}

The Bargmann space $\mathcal B^N$ with the scalar product \eqref{144} is naturally isomorphic to the Hilbert space 
$\mathcal H^N_{\rm LLL} \subset L^2(\mathbb C)^{\otimes_s N}, \mathrm d^{2N}z)$ 
consisting  of wave functions of the form 
\beq \Psi(z_1,\dots,z_N)=\psi(z_1,\dots,z_N)\exp(-\sum_j|z_j|^2/2), \qquad \psi\in\mathcal B_N\eeq with the standard $L^2$ scalar product. The energy can accordingly be  considered as a functional on this space, 
\beq \mathcal E[\Psi]=\int V_{\omega,k}(z)\rho_\Psi(z)+\langle\Psi,\mathcal I_N\Psi\rangle,\eeq
where $\rho_\Psi$ is the  one-particle density of $\Psi$ with the  normalization 
$\int \rho_\Psi(z)\,\mathrm d^2z=N$
and the potential is
\begin{equation}
\pot (z) = \om\, |z|^2 + k\,|z|^4. 
\end{equation}
Note that now $\omega<0$ is allowed, provided $k>0$. 

We shall call states with vanishing interaction energy, i.e., $\Psi\in \mathrm{ker}\, \mathcal I_N$ {\em fully correlated}, because the particles stay away from each other in the sense that the wave function vanishes if $z_i=z_j$ for some pair $i\neq j$, in sharp contrast to a fully {\it un}correlated Hartree state. The fully correlated states  in $\mathcal H^N_{\rm LLL}$ are of the form
\beq \Psi(z_1,...z_N)=\phi(z_1,\dots,z_N)\Psi_{\rm Laugh}(z_1,...z_N)\eeq 
with $\phi$ symmetric and analytic, and the Laughlin state
\beq \Psi_{\rm Laugh}(z_1,...z_N)= c \prod_{i<j} \left( z_i -z_j\right) ^2 e ^{-\sum_{j=1} ^N |z_j| ^2 / 2 }.\eeq

For an intuitive picture of the Laughlin state the following analogy may be helpful. The density $|\Psi_{\rm Laugh}(z_1,...z_N)|^2$ assigns probabilities to the possible configurations of $N$ points moving in the plane. The points like to keep a distance at least of order 1 from each other because the factors $|z_i-z_j|^4$ strongly reduce the probability when the particles are close. On the other hand the damping due to the gaussian favors a tight packing of the `balls' of size $O(1)$ around the individual particles.  The motion is strongly correlated in the sense that if one ball moves, all the other have also to move in order to satisfy these constraints.\footnote{ A colony of Emperor Penguins, sticking tightly together to survive the antarctic winter, but where the individual penguins are constantly on the move exchanging places with each other, is not a bad picture to have in mind!}

For the Hamiltonian {\it without} the anharmonic addition to the potential the Laughlin state is an {\it exact} fully correlated ground state with energy 0 and 
angular momentum $L_{\rm Laugh}=N(N-1)$. 
This is {\it not} true for $k\neq 0$ because
$ \sum_{i=1}^N \mathcal L_{(i)}^2$ does not commute with $\mathcal I_N$.
Note, however, that $\mathcal L_N$ still commutes with the Hamiltonian.\smallskip

We now address the following question: Under what conditions is it possible to tune the parameters  so that the ground state $\Psi_0$ of \eqref{162} becomes fully correlated for $N\to\infty$? The following theorem, proved in \cite{RSY}, gives sufficient conditions for this to happen. In order to state it as simply as possible we shall assume the `gap conjecture' of Subsection 6.2.3. This conjecture is not really needed, however, because is possible to replace the assumed universal gap $\Delta$ by other gaps depending on the parameters, cf. Eq. (IV.5) in \cite{RSY}.\newpage
\begin{thm}[Criteria for full correlation] 
\begin{equation}
\left\Vert P_{({\rm Ker\,}\mathcal I_N)^\perp}\Psi_0 \right\Vert  \to 0  
\end{equation}
in the limit $N\to \infty$, $\om,k\to 0$ if one of the following conditions hold:
\begin{itemize}
\item $\om \geq 0$ and
$\om N ^2 + k N ^3 \ll g \: \Delta$.\vskip.5cm
\item $0 \geq \om \geq - 2 k N$ and
$N (\om ^2/{k}) + \om N ^2 + k N ^3 \ll g \: \Delta
$.\vskip .5cm
\item $\om \leq - 2 k N$, $|\omega|/k \lesssim N^2$ and
$ k N ^{3} \ll g \: \Delta$\vskip .5cm
\item $\om \leq - 2 k N$,  $|\omega|/k \gg N^2$ and 
$ |\om| N \ll g \: \Delta$\vskip.2cm
\end{itemize}
\end{thm}

Note: For $k=0$ the first item is just the sufficient condition for the passage to the Laughlin state, $\omega/g<\Delta/N^2$, 
while the other conditions are void because $\omega<0$ is only allowed if $k>0$.


The proof of the Theorem is based on the following two items
\begin{itemize}
\item A lower bound for the ground state energy at fixed angular momentum $L$:
\beq  E_0(L)\geq (\omega+3k)L+k\frac {L^2} N.\eeq
\item An upper bound for the energy of suitable trial functions.
\end{itemize}
\medskip

The first bound is quite simple; it follows essentially from
\beq \sum_i \mathcal L_{(i)}^2\geq \frac 1N\left (\sum_i \mathcal L_{(i)}\right)^2\eeq
 that holds because $\mathcal L_{(i)}$ and $\mathcal L_{(j)}$ commute for any $i,j$.


The upper bound is achieved by means of  trial states of the form `giant vortex times Laughlin', namely,
with $m\geq 0$ and $c_{m,N}$ a normalization constant,
\beq  \Psi_{\rm gv}^{(m)}(z_1,\dots,z_N) = c_{m,N} \prod_{j=1} ^N z_j ^m \prod_{i<j} \left( z_i - z_j\right) ^2 e^{-\sum_{j=1} ^N |z_j| ^2 / 2} \eeq 
For small $m$ these are Laughlin's `quasi hole' states \cite{Lau}  but for $m\gtrsim N$, i.e., $mN\gtrsim N^2$= angular momentum of the Laughlin state, the label `giant vortex' appears more appropriate. Note, however, that mathematically and physically these states are rather different from the previously considered {\it uncorrelated} giant vortex states in Section 5.6.2.



The energy of the trial states can be estimated using properties of the angular momentum operators and the radial symmetry in each variable of $\prod_{j=1} ^N |z_j| ^{2m}\times$ the gaussian measure. Optimizing the estimate over $m$ leads to

\begin{equation}\label{eq:intro m opt}
m_{\rm opt} = \begin{cases}
               0& \mbox{\rm if } \om \geq - 2 k N \\
               \frac{|\om|}{2 k}- N &\mbox{\rm if } \om < - 2 k N.
              \end{cases} 
\end{equation}
\medskip

This is consistent with the picture that the Laughlin state is an approximate ground state in the first two cases of Theorem 1, in particular for negative $\omega$ as long as  $|\omega|/k\lesssim N$. The angular momentum remains $O(N^2)$ in these cases.



When $\omega<0$ and  $|\omega|/(kN)$ becomes large  the angular momentum is approximately  $L_{\rm qh}=O(N|\omega|/k)\gg N^2$, much larger than for the Laughlin state.
A further transition  at $|\omega|/k\sim N^2$ is manifest through the change of the subleading contribution to the energy of the trial functions. Its order of magnitude changes from $O(kN ^3)$ to $O(|\om| N)$ at the transition.\smallskip

To obtain further insights into the physics of the transition we consider the density of the trial wave functions. This analysis  \cite{RSY2} is based on the analogy of the 
$N$-particle density with the Gibbs distribution of a 2D Coulomb gas \cite{Lau}. Taking a mean field limit of this system brings out the essential features of the single particle density for large $N$, in particular its {\it incompressibility}.

\subsection {The $N$-particle Density as a Gibbs Measure}

We denote $(z_1,\dots,z_N)$ by $Z$ for short and consider the scaled $N$ particle density (normalized to 1)
\beq \rhoNm (Z) := N ^N \left| \Psi_{\rm gv}^{(m)} (\sqrt{N} Z )\right| ^2.\eeq
We can write
\begin{eqnarray*}\hskip-.1cm
\rhoNm (Z) &=& \mathcal Z_{N,m}^{-1} \exp\left( \sum_{j=1} ^N \left( - N  |z_j| ^2 + 2 m \log |z_j|\right)  + 4 \sum_{i<j} \log |z_i - z_j|\right) 
\\  &=& \mathcal Z_{N,m}^{-1} \exp\left( -\frac{1}{T} \HNm(Z) \right),
\end{eqnarray*}
with $T=N^{-1}$ and 
\beq \HNm(Z)=\sum_{j=1} ^N \left(   |z_j| ^2 - \frac{2 m}N \log |z_j|\right)  - \frac 4N \sum_{i<j} \log |z_i - z_j|.\eeq


\subsubsection {Plasma analogy and mean field limit}

The Hamilton function $\HNm(Z)$ defines a classical 2D Coulomb gas (`plasma', `jellium') in a uniform  background of opposite charge and with a point charge $(2m/N)$ at the origin, corresponding respectively to the $|z_i|^2$ and the $-\frac{2 m}N \log |z_j|$ terms. \medskip

The probability measure $\rhoNm (Z)$ minimizes the free energy functional
\beq \mathcal F(\rho)=\int_{\mathbb R^{2N}} \HNm(Z) \rho(Z)+T\int_{\mathbb R^{2N}}\rho(Z)\log\rho(Z)\eeq
for this Hamiltonian at $T=N^{-1}$. \medskip

The $N\to\infty$ limit is in this interpretation a {mean field limit} where at the same time $T\to 0$. It is thus not unreasonable to expect that for large $N$, and in an appropriate sense,
\beq \rhoNm\approx \rho^{\otimes N}\eeq
with a {one-particle} density $\rho$ minimizing a {mean field free energy functional.}



The {mean field free energy functional} is defined as
\begin{equation}
\MFmf [\rho]: = \intR  W_m\,  \rho - 2 \intR\intR \rho(z)\log|z-z'|\rho(z') + N ^{-1} \int_{\R ^2} \rho \log \rho 
\end{equation}
with
\begin{equation}
W_m (z) = |z| ^2 - 2 \frac{m}{N} \log |z|.
\end{equation}
It has a minimizer $\rhoMFm$ among probability measures on $\mathbb R^2$ and this minimizer is in \cite{RSY2}  {proved} to be a good approximation for the scaled {1-particle density} of the trial wave function, i.e., 
$$\rhoNmone (z):= \int_{\R ^{2(N-1)}} \rhoNm (z,z_2,\ldots,z_N) \mathrm d^2z_2\ldots \mathrm d^2z_N.$$
(Recall the scaling: This density in the scaled variables $z$ is normalizes so that its integral is 1. The corresponding density in the physical, unscaled variables $\zeta=\sqrt N z$ has total mass $N$.)

\subsubsection{Asymptotic formulas for the mean field density}

The picture of the 1-particle density that arises from asymptotic formulas for the mean-field density is as follows:\medskip

If $m\leq N^2$, then  $\rho^{\rm MF}_m$ is well approximated by a density $\hat \rho^{\rm MF}_m$ that minimizes the mean field functional without the entropy term.\medskip

 It takes a {\em constant value}\footnote{Note that this is a statement about the mean field density that is a good approximation in a weak sense (but not pointwise) to the true 1-particle density for large $N$. See \cite{Ciftja} for numerical calculations of the true density for $N=400$.}  $(2\pi)^{-1}$ (for all $N$)on an annulus with inner and outer radii (in the scaled variables!)
$$R_-=(m/N)^{1/2},\qquad R_+=(2+m/N)^{1/2}$$ and is zero otherwise.  The constant value is a manifestation of the {\it incompressibility of the density} of the trial state.
\medskip

For $m\gtrsim  N^2$ the entropy term dominates the interaction term\newline $\int\int\rho(z)\log|z-z'|\rho(z')$. The density is well approximated by the {Gaussian} $\rho^{\rm th}(z)\sim |z|^{2m}\exp(-N|z|^2)$ that is centered around $(m/N)^{-1/2}$ but has maximal value $\sim N/m^{1/2}\ll 1$ for $m\gg N^2$.



As the parameters $\omega$ and $k$ tend to zero and $N$ is large the qualitative properties of the optimal trial wave functions thus exhibit different phases:\medskip
\begin{itemize}
\item The  state changes from a pure Laughlin state to a modified Laughlin state with a `hole' in the density around the center when $\omega$ is negative and $|\omega|$ exceeds $2kN$.\medskip

\item A further transition is indicated at $|\omega|\sim kN^2$. The density profile changes from being `flat' to a Gaussian.
\end{itemize}


\begin{figure}[htf]
\center
\fbox{\includegraphics[width=12cm, bb=100 450 500 780]{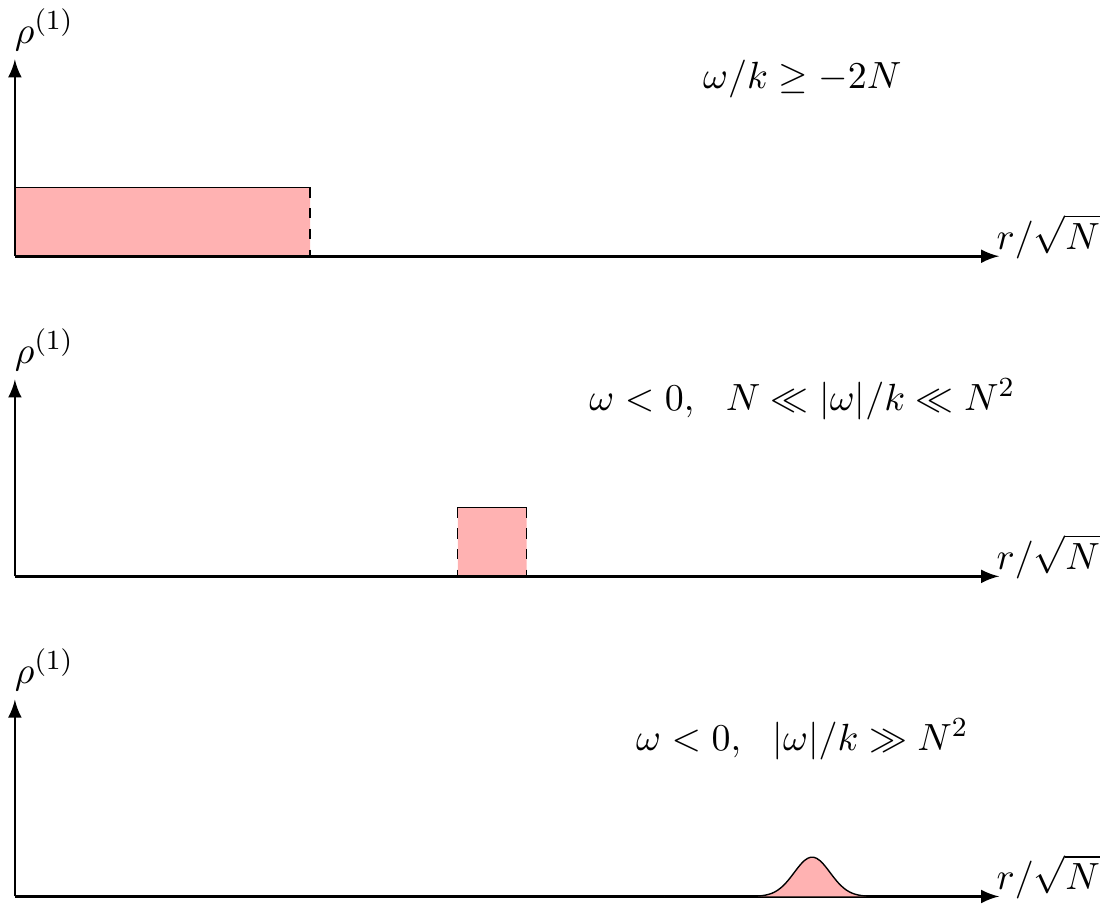}}
\caption{The three phases of the density $\rho^{(1)}_{N,m}$ (not to scale).}
\label{fig5}
\end{figure}

An intuitive understanding of these transitions may be obtained by employing the previous picture of the points $z_i$ as being the centers of essentially non overlapping balls of size $O(1)$. From the point of view of the plasma analogy the particles stay away from each other because of the repulsive Coulomb potential between them, while the attractive external potential due to the uniformly charged background keeps them as close together as possible. Modifying the wave function by a factor $\prod_j z_j^m$ has the effect of a repulsive charge of magnitude $m$ at the origin that pushes the particles (collectively!) away from the origin, creating a `hole'. The effect of such a hole on the energy of the wave function in the trap potential is to increase the energy if $\omega$ is positive. Hence the ground state will not have a hole. If $\omega<0$ the effective trapping potential has a Mexican hat shape with a minimum away from the origin, but since no ball can move without `pushing' all the other balls, it is too costly for the system to take advantage of this as long as $k$ stays above the critical value $|\omega|/2N$. For smaller $k$ a hole is formed. The balls remain densely packed until the minimum of the Mexican hat potential moves so far from the origin that an annulus of width $O(1)$ at the radius of the minimum can accommodate all $N$ balls. This happens for $k\lesssim  |\omega|/N^2$. For smaller $k$ (larger radius) the balls need not be tightly packed in the annulus and the average local density decreases accordingly.

\subsection{Summary and Conclusions}

The main conclusion from the analysis presented above of of many-body ground states in the lowest Landau level generated by fast rotation can be summarized as rolls:

\begin{itemize}
\item The parameter regime $g\ll N\omega$, i.e., $L\ll N^2$, can be described by a GP theory in the LLL.
\item To enter the `fully correlated' regime with $L\geq N(N-1)$ we have studied a rotating Bose gas in a quadratic plus quartic trap (coupling $k$) where the rotational frequency can exceed the frequency of the quadratic part of the trap, i.e, the frequency difference $\omega$ can be  negative. 

\item Through the analysis of  trial states for energy upper bounds and simple lower bounds we have obtained criteria for the ground state to be fully correlated in an asymptotic limit. The lower bounds, although not sharp, are of the same order of magnitude as the upper bounds.

\item The density of the wave functions can be analyzed through the plasma analogy. The character of the density changes at $|\omega|/k=O(N)$ and again at $|\omega|/k=O(N^2)$.

\end{itemize}


\begin{acknowledgement}
I thank Nicolas Rougerie for valuable comments, Christian K\"ohler, Matthias Plaschke, Mathieu Lewin and Robert Seiringer for help with the figures, 
and the Austrian Science Fund (FWF) for support under Project P 22929-N16. 
\end{acknowledgement}

\end{document}